\documentclass{osa-article}

\journal{osajournal}

\articletype{Research Article}

\usepackage{siunitx}

\usepackage{multicol}

\usepackage[normalem]{ulem}
\usepackage{lineno}

\newcommand{\mmr}{\mathrm{m}}
\newcommand{\eff}{\mathrm{eff}}
\newcommand{\tot}{\mathrm{tot}}

\newcommand{\zpf}{\mathrm{zpf}}
\newcommand{\xzp}{\mathrm{xzp}}

\newcommand{\fbmr}{\mathrm{fb}}
\newcommand{\auxmr}{\mathrm{aux}}
\newcommand{\qmr}{\mathrm{q}}
\newcommand{\impmr}{\mathrm{imp}}

\newcommand{\dd}{\mathrm{d}}

\begin{document}

\title{Laser cooling a membrane-in-the-middle system close to the quantum ground state from room temperature}

\author{Sampo~A.~Saarinen,\authormark{1,2,*} Nenad Kralj,\authormark{1,2,3,*} Eric~C.~Langman,\authormark{1,2} Yeghishe Tsaturyan\authormark{1,2,4} and Albert Schliesser\authormark{1,2,$\dagger$}}

\address{
\authormark{1} Niels Bohr Institute, University of Copenhagen, Blegdamsvej 17, 2100, Copenhagen, Denmark\\
\authormark{2} Center for Hybrid Quantum Networks, Niels Bohr Institute, University of Copenhagen, Blegdamsvej 17, 2100, Copenhagen, Denmark\\
\authormark{3} Institute for Gravitational Physics, Leibniz Universit\"{a}t Hannover, Callinstraße 38, 30167 Hannover, Germany\\
\authormark{4}Present address: Pritzker School of Molecular Engineering, University of Chicago, 5640 South Ellis Avenue, Chicago, Illinois 60637, USA\\
\authormark{*}These authors contributed equally: Sampo A. Saarinen, Nenad Kralj}

\email{\authormark{$\dagger$}albert.schliesser@nbi.ku.dk} 



\begin{abstract}
Many protocols in quantum science and technology require initializing a system in a
pure quantum state.
In the context of the motional state of massive resonators, this enables studying fundamental physics at the elusive quantum-classical transition, and measuring force and acceleration with enhanced sensitivity. 
Laser cooling has been a method of choice to prepare mechanical resonators in the quantum ground state, one of the simplest pure states. 
However, in order to overcome the heating and decoherence by the thermal bath, this usually has to be combined with cryogenic cooling.
Here, we laser-cool an ultracoherent, soft-clamped mechanical resonator close to the quantum ground state directly from room temperature. 
To this end, we implement the versatile membrane-in-the-middle setup with one fiber mirror and one phononic crystal mirror, which reaches a quantum cooperativity close to unity already at room temperature.
We furthermore introduce a powerful combination of coherent and measurement-based quantum control techniques, which allows us to mitigate thermal intermodulation noise. 
%
The lowest occupancy we reach is {30} phonons, limited by measurement imprecision.
Doing away with the necessity for cryogenic cooling should further facilitate the spread of optomechanical quantum technologies.
%
\end{abstract}


\section{Introduction}
Over the last decade, the relative simplicity and great versatility of membrane-in-the-middle (MIM) systems \cite{Thompson2008} have rendered them a highly popular choice for a variety of optomechanics experiments.
To name just a few examples, they have been used for early demonstrations of optomechanical effects such as dynamical \cite{Wilson2009, Jayich2012, FlowersJacobs2012, Rochau2021} and quantum backaction \cite{Purdy2013}, optomechanically induced transparency \cite{Karuza2013} and Raman ratio thermometry \cite{Purdy2015,Galinskiy2020},
to implement coupling between mechanical and atomic degrees of freedom \cite{Moller2017,Christoph2018,Karg2020,Thomas2021},
to explore novel routes for enhanced sensitivity in gravitational wave detection \cite{Ma2014,Qin2014,Page2021},
to study multimode \cite{Shkarin2014, Nielsen2017, Fedoseev2021} and topological physics \cite{Xu2016},
to manipulate the quantum state of light \cite{Purdy2013Squeeze, Chen2020},
and to realize electro-opto-mechanical transducers \cite{Bagci2014,Andrews2014, Higginbotham2018, Brubaker2021}.
More recently, force sensing \cite{Fischer2019, Kosata2020,Catalini2020} and imaging \cite{Halg2021} applications have received particular attention, due to the very low thermomechanical force noise \cite{Reinhardt2016, Norte2016, Tsaturyan2017}, and the potential for very efficient optical detection of the sensor state \cite{Rossi2018,Rossi2019}. 
Indeed, using a broadband correlation technique \cite{Vyatchanin1995,Kampel2017}, this platform has even allowed continuous force measurements beyond the standard quantum limit (SQL) for the first time \cite{Mason2019}.

However, in order to reach the quantum regime, it has so far been necessary to cool the membrane to cryogenic temperatures \cite{Purdy2013, Purdy2015, Galinskiy2020,Moller2017, Christoph2018, Thomas2021, Nielsen2017, Purdy2013Squeeze, Chen2020, Andrews2014, Higginbotham2018, Brubaker2021, Rossi2018, Rossi2019, Mason2019}.
For even greater flexibility and versatility, it would be
desirable to be able to operate in the quantum regime already at room temperature.
To date,
few mechanical systems have even approached the quantum regime at room temperature, including optically levitated nanoparticles~\cite{Delic2020,Magrini2021},  select nano- \cite{Guo2019,Sudhir2017,Purdy2017} and micromechanical devices \cite{Cripe2019},  as well as a set of macroscopic mirrors in a gravitational wave observatory \cite{LIGO2021}.
Save for the latter, a broad range of applications have yet to be demonstrated with these platforms, in contrast to the MIM systems.
Our goal is therefore to implement a room-temperature MIM system in the quantum regime. 

A key enabler towards this goal is a recently developed ultracoherent mechanical system, a ``soft-clamped'' membrane resonator, whose coherence \emph {at room temperature} rivals those of standard membranes at cryogenic temperatures \cite{Purdy2013,Nielsen2017}.
We also bring to bear several techniques to mitigate background thermal noises in the system, including a phononic mirror and
a combination of  sideband cooling and feedback cooling.

\section{Soft-clamped membrane-in-the-middle system}
\begin{figure}[htb]
    \centering
    \includegraphics[width=0.83\textwidth]{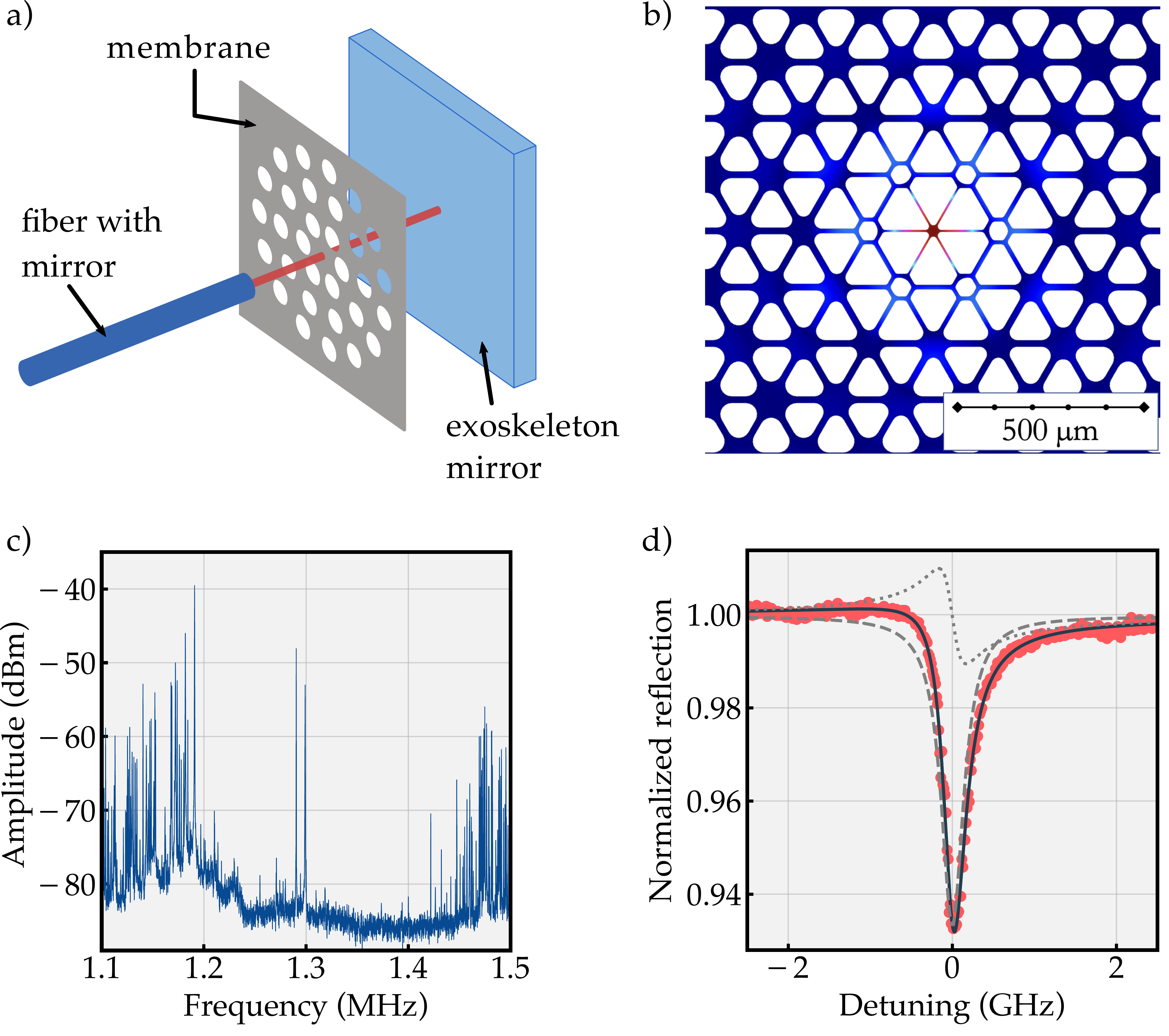}
    \caption{a)~Sketch of the cavity assembly with main components. b)~Top view of membrane geometry, with a central defect in a phononic crystal. Color code indicates simulated out-of-plane displacement (increasing from blue to red shades). c)~Thermal spectrum of the dandelion, as measured inside the fiber cavity. It features a bandgap in the region from \SIrange{1.2}{1.5}{\mega\hertz} and two soft-clamped defect modes around its center. d)~Cavity resonance as measured in reflection. The asymmetric lineshape is a consequence of the filtering done by the fiber mode. The fit to the model (solid black line) can be decomposed into the Lorentzian and dispersive contributions (dashed and dotted light-gray lines, respectively).}
    \label{fig:fig1}
\end{figure}

The membrane is placed  inside a Fabry-P\'{e}rot optical resonator (Fig.~\ref{fig:fig1}), between a node and an anti-node of the standing optical wave in the cavity.
We thereby implement the standard MIM dispersive optomechanical coupling \cite{Thompson2008} described by the Hamiltonian \cite{Aspelmeyer2014, Bowen2016} $H_\mathrm{om}=\hbar g_0 a^\dagger a (b+b^\dagger)$. 
(Here, $\hbar$ is the reduced Planck constant, $g_0$ the vacuum optomechanical coupling rate, and $a(a^\dagger)$ and $b(b^\dagger)$ are the optical and mechanical annihilation (creation) operators, respectively.)
The membrane is 15 nm thick and patterned with a phononic crystal, which affords soft-clamping of an out-of-plane mechanical mode localized close to an intentionally introduced crystal defect \cite{Tsaturyan2017}.
In this work, we introduce a new design for both the crystal and the defect (which we refer to as the `Dandelion' design), see {Fig.~\ref{fig:fig1}b)}.
Compared to the design used in our previous works~\cite{Tsaturyan2017,Rossi2018, Rossi2019, Mason2019,Chen2020,Thomas2021,Catalini2020}, Dandelion-class defects exhibit significantly lower masses (up to a 30-fold reduction) when fabricated from identical substrates and targeting the same frequency and $Q$-factors. 
Additionally, the mechanical mode can be centered in an $\sim$\SI{50}{\percent} bigger bandgap, with typically only one extra mechanical mode in the bandgap (2 total). However, these benefits come with a significant reduction in the size of the pad at the defect center, which can lead to complications with large cavity
waists and/or optical alignment. As such, the optimal Dandelion for any application is not necessarily the design with the smallest mass, highest
$Q$-factor, or other figure of merit. 

Specifically due to alignment tolerances, the particular design used in this work is a medium-size Dandelion with a central pad approximately \SI{30}{\micro\meter} across, resulting in a mechanical mode with resonance frequency $\Omega_\mmr/2 \pi = \SI{1.3}{\mega\hertz}$ and modal mass $m_\mathrm{eff}\approx \SI{200}{\pico\gram}$. 
%
This results in a relatively large amplitude of zero-point fluctuations, $x_\mathrm{zpf}=\sqrt{\hbar/ 2 m_\eff \Omega_\mmr} =5{.}7\,\mathrm{fm}$.
Due to the soft clamping effect, we achieve intrinsic damping rates $\Gamma_\mathrm{m}/2\pi\approx \SI{9}{\milli\hertz}$, corresponding to quality factors as high as $Q_\mmr = \Omega_\mmr/\Gamma_\mmr=\num{1.4e8}$ with these devices at room temperature, $T\approx \SI{300}{\kelvin}$, where the occupation of the mechanical bath is as high as $\bar n_\mathrm{th}\approx k_\mathrm{B} T/\hbar \Omega_\mmr\approx 5{.}1\cdot 10^6$ ($k_\mathrm{B}$ is the Boltzmann constant).
This implies a thermomechanical decoherence rate of $\gamma= \bar n_\mathrm{th} \Gamma_\mmr \approx 2 \pi \cdot \SI{48}{\kilo\hertz}$.
In turn, this corresponds to $\Omega_\mmr/\gamma\approx 27$ coherent oscillations, comparing favourably to similar experiments (four coherent oscillations \cite{Guo2019}) and even levitated optomechanical systems ($\lesssim 15$ coherent oscillations \cite{Delic2020}).

The cavity consists of one curved and one flat mirror.
We fabricate the curved mirror by laser-machining a concave
indentation
into the facet of a single-mode fiber, and subsequently coating it with a high-reflectivity coating \cite{Hunger2010,Hunger2012}.
To realize the flat mirror, a slightly different coating is deposited on a flat pyrex glass wafer that has been bonded to a silicon ``exoskeleton'' support structure, whose role will be detailed in section \ref{ssec:excessNoise}.
The mirrors are set up to face each other with a distance of $L \approx \SI{95}{\micro\meter}$.
Details on the fabrication of the mirrors and the mechanical setup are provided in
Supplement~1.

The mirror coatings are designed to provide a one-sided, high finesse cavity at a wavelength of around \SI{1550}{\nano\meter}.
Independent measurements of the actually realized mirrors' power transmissivities, $T_f$ and $T_e$ for the fiber and exoskeleton mirror, respectively, suggest an overcoupling of 
$\eta_\mathrm{c} = T_f/(T_f+T_e) \approx 0.9$ for the cavity coupled through the fiber mirror.
For the assembled cavity including a membrane,
we measure the reflection signal through the (more transmissive) fiber mirror, as shown in Fig.~\ref{fig:fig1}d). 
Using the model of Ref.~\cite{Gallego2016} to fit the data, we obtain a FWHM cavity linewidth of $\kappa/2\pi = \SI{340}{\mega\hertz}$, corresponding to a finesse of $\mathcal{F}\approx 4400$.
%
Note that the spatial filtering of the cavity leakage by the fiber mode leads to an asymmetric reflection signature (with two contributions of Lorentzian and dispersive shape respectively), the depth of which is not limited by the mode-matching efficiency $\varepsilon$ between the cavity and detected optical modes.
See Supplement~1 for more details on the model.
%

 The geometry of the
 indentation
 in the fiber mirror is approximately spherical with a \SI{300}{\micro\meter}-radius of curvature, leading to an approximately $\SI{10}{\micro\meter}$ waist of the cavity's fundamental optical mode.
 This allows good transverse overlap ($\xi\sim 1$) of the optical and mechanical mode shapes, even for the small---and therefore low-mass---defect of the Dandelion design.
 Together with the short cavity length, this facilitates large vacuum optomechanical coupling, which can, for the optimal position of the membrane within the cavity standing wave, be approximated as
 \begin{equation}
     \label{eq:g0}
     g_0^\mathrm{max}\approx 2 \frac{\omega_\mathrm{c}}{L} \, |r| \, x_\mathrm{zpf} \, \xi,
 \end{equation}
 where $\omega_\mathrm{c}$ is the cavity resonance frequency and $r$ is the membrane's optical (field) reflectivity.
 Indeed, for the parameters of the system, we expect a coupling rate of $g_0/2 \pi\lesssim 3{.}3\,\mathrm{kHz}$.
 %
%

\section{Dynamical backaction}
\label{s:approach}

Broadly speaking, to reach the quantum regime, it is necessary to achieve a quantum cooperativity $C_\qmr = 4 g^2/(\kappa \gamma) \gtrsim 1$, where $\kappa$ and $\gamma$ are the optical and mechanical decoherence rates, and $g= g_0 \sqrt{\bar n_{\mathrm{cav}}}$ is the field-enhanced coupling rate with the intracavity photon number $\bar n_{\mathrm{cav}}$.
In a first set of experiments, we infer the obtained coupling by analysing dynamical backaction \cite{Wilson2009, Jayich2012, FlowersJacobs2012, Rochau2021} on the mechanical mode.

\begin{figure}[hbt]
    \centering
    \includegraphics[width=0.55\columnwidth]{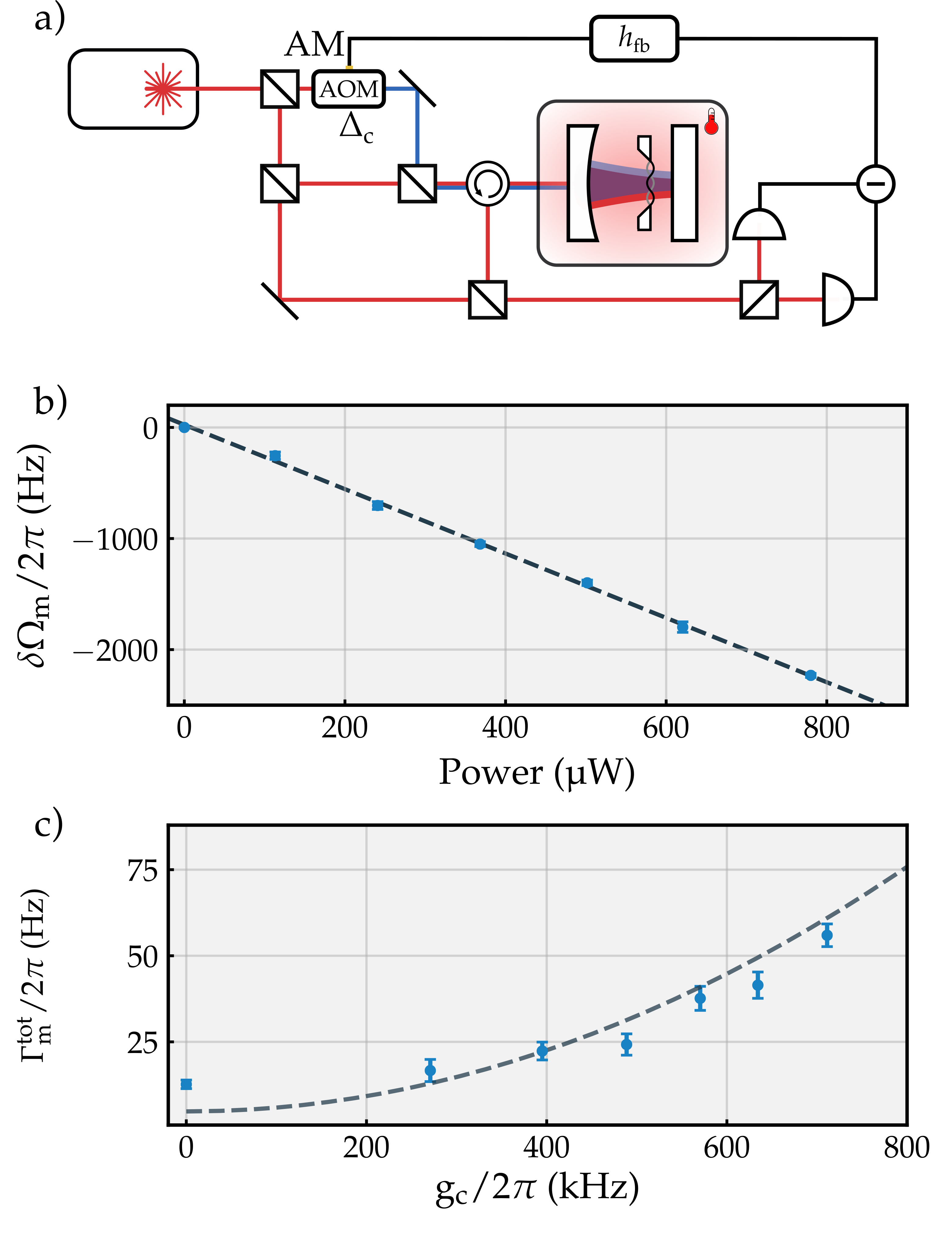}
    \caption{a)~Schematic of the experimental setup. The probe (red) and cooling (blue) beams are derived from a single laser. The probe is locked on resonance with the cavity and used for a homodyne readout of the membrane motion. The feedback force is exerted by convolving this measurement with a filter function $h_\fbmr$ and using it to modulate the amplitude of the cooling beam. The cooling beam is detuned from the cavity by $\Delta_\mathrm{c}/2\pi = \SI{-80}{\mega\hertz}$ and thus also provides sideband cooling. The optomechanical cavity is kept in a vacuum chamber at a pressure $p_\mathrm{UHV} \approx \SI{2e-8}{\milli\bar}$. b)~The mechanical frequency shift due to the cooling beam as a function of input optical power. The dashed line is a linear fit to the data following eq.~(\ref{eq:opt_spring}), with $g_\mathrm{c}^2 \propto (\mathrm{Power})$. c)~The optically broadened mechanical linewidth as a function of the cooling beam coupling rate $g_\mathrm{c}$. With $g_\mathrm{c}$ extracted from the measured power based on the fit in b), 
    the only fit parameter for the dashed line is a vertical offset due to a small cooling by the probe.
    The data points in both b) and c) are mean values over five acquisitions and the error bars correspond to the standard errors of the mean.}
    \label{fig:Fig_2}
\end{figure}

To this end, we couple a low-noise laser (NKT Photonics Koheras Basik E15) of wavelength $\lambda \approx  1542\,\mathrm{nm}$ to the cavity.
We
derive two beams from this laser: a ``probe'' beam to lock the cavity on resonance with one of its fundamental modes using a Pound-Drever-Hall locking scheme, and to read out the mechanical motion via a homodyne interferometer.
%
%
%
A second, ``cooling'' beam is offset by \SI{-80}{\mega\hertz} via a fiber-based acousto-optic modulator and its polarization is made orthogonal to that of the probe. These two facts help minimize the unwanted effect of the cooling beam on the measurement record. 
Figure~\ref{fig:Fig_2}a) shows the schematic of the experimental setup, an extended discussion of which is presented in
Supplement~1.

As we increase the cooling beam power, we observe the evolution of the mechanical noise spectrum through the probe homodyne detector. 
We extract the mechanical frequency shift 
$\delta\Omega_\mmr^\mathrm{c}$
resulting from the optical spring effect, as a function of the cooling beam power (cf. Figure~\ref{fig:Fig_2}b)).
The standard model for the mechanical frequency shift by an optical spring~\cite{Aspelmeyer2014},
\begin{equation}
    \delta\Omega_\mathrm{m}^\mathrm{c}=
    g_\mathrm{c}^2 \left(
    \frac{\Delta_\mathrm{c}-\Omega_\mmr}{\kappa^2/4+(\Delta_\mathrm{c}-\Omega_\mmr)^2}+
    \frac{\Delta_\mathrm{c}+\Omega_\mmr}{\kappa^2/4+(\Delta_\mathrm{c}+\Omega_\mmr)^2}
    \right) \, ,
    \label{eq:opt_spring}
\end{equation}
supplied with the independently known
$\Omega_\mmr$,
$\kappa$ and detuning $\Delta_\mathrm{c}$
, allows us to extract the field-enhanced coupling rate $g_\mathrm{c}$ as a function of measured power.
%
The index ``c'' here refers to parameters of the cooling beam.
For a given power, we can estimate the intracavity photon number by estimating the incoupling efficiency of light.
%
In this manner, we infer an experimental value for the vacuum optomechanical coupling rate of
$g_0/2\pi = \SI{2.3}{\kilo\hertz}$,
in good agreement with the theoretical value, assuming slightly sub-optimal longitudinal and transverse positioning of the membrane.

We also compare the observed broadening of the mechanical linewidth
with the expected damping due to dynamical backaction of the cooling beam~\cite{Aspelmeyer2014},
\begin{equation}
    \Gamma_\mathrm{m}^\mathrm{c}=g_\mathrm{c}^2 \left(
    \frac{\kappa}{\kappa^2/4+(\Delta_\mathrm{c}+\Omega_\mmr)^2}-
    \frac{\kappa}{\kappa^2/4+(\Delta_\mathrm{c}-\Omega_\mmr)^2}
    \right) \, ,
\end{equation}
for all of the
inferred couplings $g_\mathrm{c}$, and find good agreement.
Specifically, in Figure~\ref{fig:Fig_2}c) we show the total damping rate $\Gamma_\mathrm{m}^\mathrm{tot}=\Gamma_\mathrm{m}+\Gamma_\mathrm{m}^\mathrm{p}+\Gamma_\mathrm{m}^\mathrm{c}$, with
$\Gamma_\mathrm{m}^\mathrm{p}$
the contribution from optical damping by the (very slightly red-detuned) probe beam.
To allow for this probe-induced damping, the dashed line is a fit with the vertical offset as the only free parameter, but is otherwise fully constrained by independently known parameters ($\Omega_\mmr$, $\kappa$ and $\Delta_\mathrm{c}$) and the fit of the optical spring. 
The fit to the linewidths yields $\Gamma_\mathrm{m}^\mathrm{p}/2\pi = \SI{4.6}{\hertz}$, corresponding to a probe detuning $\Delta_\mathrm{p}/2\pi \lesssim \SI{-2}{\mega\hertz}$. The zero-power linewidth is likely overestimated
due to inhomogeneous broadening caused by a fluctuation in coupled laser power.


Finally, we confirm that we can increase the probe power further, into a regime where  $C_\mathrm{q}\approx 1$ would be expected, without the onset of optomechanical (or other) instabilities.

\section{Laser cooling close to the quantum ground state}
\label{sec:feedback}
\subsection{Theoretical cooling limits}
Next we aim at preparing the mechanical system in a pure quantum state---specifically, the quantum ground state.
That is, starting from the equilibrium value $\bar n_\mathrm{th}=5{.}1\cdot 10^6$, we want to reduce the residual mechanical occupation $\bar n=\langle b^\dagger b\rangle\rightarrow 0$.
Laser sideband cooling has been used with great success to cool mechanical oscillators, including MIM systems \cite{Wilson2009, Jayich2012, FlowersJacobs2012, Nielsen2017, Mason2019, Rochau2021}.
Adapting the theory of sideband cooling yields a final occupation \cite{Aspelmeyer2014}
\begin{equation}
   \bar n=\frac{ \bar n_\mathrm{th} \Gamma_\mmr (1+C_\mathrm{q}) + \bar n_\mathrm{m}^\mathrm{c} \Gamma_\mathrm{m}^\mathrm{c}}{\Gamma_\mathrm{m}^\mathrm{tot} } \, ,
   \label{e:barn}
\end{equation}
where
$\Gamma_\mmr^\mathrm{tot}$ in the denominator allows for a small probe detuning, the numerator accounts for the probe's quantum backaction (with $C_\mathrm{q}\lesssim 1$ the probe's quantum cooperativity) and
\begin{equation}
   \label{e:barnmin}
  \bar n_\mathrm{m}^\mathrm{c}=\frac{(\Omega_\mmr+\Delta_\mathrm{c})^2+(\kappa/2)^2}{-4 \Delta_\mathrm{c} \Omega_\mmr}
\end{equation}
is the minimum occupancy that can be reached in the limit of strong cooling $\Gamma_\mathrm{m}^\mathrm{c}\gg \bar n_\mathrm{th} \Gamma_\mmr(1+C_\mathrm{q})$, due to quantum backaction of the cooling beam. 
In the unresolved sideband regime with $\kappa \gg \Omega_\mmr$, in which we work here, $\bar n_\mathrm{m}^\mathrm{c}\approx \kappa/4\Omega_\mmr$  for optimal detuning $\Delta_\mathrm{c}$, corresponding to about 65 phonons for the parameters of our system.
We note that eq.~(\ref{e:barn}) should, in principle, also contain classical backaction, i.e. photothermal heating and intensity noise heating, which would manifest as an effective thermal bath occupation that increases with optical power.
The analysis of the mechanical area as a function of optical power presented in Supplement~1 shows, however, that this effect is negligible.

Given that sideband cooling to the ground state is prohibited by quantum backaction, we additionally
expect to be using measurement-based quantum control techniques \cite{Mancini1998, Cohadon1999, Kleckner2006,Bowen2016, Wilson2015, Rossi2018, Rossi2019, Magrini2021}
in order to reach it.
In these approaches, one tracks the mechanical position $x(t) = x_\zpf(b(t) + b^\dagger(t))$ through the optical signal, in our case the output $y(t)$ of the homodyne interferometer detecting the probe light.
%
A sufficiently precise measurement record allows tracking the state of the oscillator, and preparing it in a highly pure conditional quantum state~\cite{Rossi2019}. 
Alternatively, one can derive a feedback force $F_\fbmr(t) = h_\fbmr(t) \ast y(t)$ from the measured displacement, where  $h_\fbmr(t)$ is a filter kernel and `$*$' denotes a convolution.
Applying this force to the mechanical resonator allows preparing it in its ground state unconditionally \cite{Rossi2018}.

Importantly, the measured signal  $y(t) = x(t) + x_\impmr(t) $
contains, apart from the true  position $x(t)$, also the imprecision $x_\impmr(t)$, a noise background inevitable in the kind of interferometric measurement discussed here \cite{Braginsky1992}.
For the optimal gain in the feedback loop, the minimum occupation that can be reached can be written approximately as (see Supplement~1)
\begin{equation}
    \label{eq:nMinFbBasic}
    \bar n_\mathrm{min}^\mathrm{fb} \approx \sqrt{\frac{\bar S_{FF}^\tot \cdot \bar S_{xx}^\impmr }{4 \hbar^2 }}-\frac{1}{2} \, .
\end{equation}
Here, $\bar S_{FF}^\tot$ is the (symmetrized, single-sided) spectral density of force fluctuations driving the mechanical resonator, whereas $\bar S_{xx}^\impmr$ is the spectral density of $x_\impmr$.
We assume these spectral densities are constant in the relevant range around the mechanical frequency $\Omega_\mmr$ and have dropped the frequency argument for brevity.
The total force fluctuations $\bar S_{FF}^\tot =\bar S_{FF}^\mathrm{th}+\bar S_{FF}^\mathrm{rp}$ contain both the thermal Langevin force noise $\bar S_{FF}^\mathrm{th}=4 m_\eff \Gamma_\mmr k_\mathrm{B} T$, as well as the radiation pressure fluctuations of the intracavity light field(s) $\bar S_{FF}^\mathrm{rp}$.
If the latter are given only by quantum fluctuations, we have $\bar S_{FF}^\mathrm{rp}=C_\mathrm{q} \bar S_{FF}^\mathrm{th}$.

The imprecision, in turn, must at least contain the quantum (shot) noise in the optical detector.
This ensures that the Heisenberg measurement-disturbance uncertainty relation $\bar S_{FF}^\mathrm{rp} \bar S_{xx}^\impmr \geq \hbar^2$ is always respected \cite{Braginsky1992,Aspelmeyer2014,Bowen2016}.
For imperfect detection efficiency $\eta_\mathrm{det}<1$, the imprecision as referenced to equivalent mechanical displacement grows, and $\bar S_{FF}^\mathrm{rp} \bar S_{xx}^\impmr = \hbar^2/ \eta_\mathrm{det}$ if both radiation pressure and imprecision noise are limited by quantum fluctuations otherwise \cite{Schliesser2010}.
We can additionally take into account the classical contribution to the imprecision noise $\bar S_{xx}^\mathrm{imp,cl}$, for example due to fluctuations of the cavity mirrors which produce cavity length changes unrelated to (but indistinguishable from) membrane motion.
We then have $\bar S_{xx}^\mathrm{imp} = \bar S_{xx}^\mathrm{imp,q} + \bar S_{xx}^\mathrm{imp,cl}$ and $\bar S_{xx}^\mathrm{imp,q} =\hbar^2/ \eta_\mathrm{det} C_\mathrm{q} \bar S_{FF}^\mathrm{th}$, with which we can rewrite eq.~(\ref{eq:nMinFbBasic}) as
\begin{equation}
   \label{eq:nMinFbAdvanced}
       \bar n_\mathrm{min}^\mathrm{fb} 
       \approx 
       \sqrt{
            \left(1+C_\mathrm{q} \right)
            \left(\frac{1}{4 \eta_\mathrm{det} C_\mathrm{q}}+\frac{\bar S_{xx}^\mathrm{imp,cl}\cdot \bar n_\mathrm{th} \Gamma_\mmr}{2 x^2_\mathrm{zpf}} \right)
       }-\frac{1}{2} \, .
\end{equation}
Evidently, for quantum-limited ($\bar S_{xx}^\mathrm{imp,cl}\rightarrow 0$), strong ($C_\mathrm{q}\gg 1$) and efficient ($\eta_\mathrm{det}\rightarrow 1$) measurement, the ground state can be reached ($\bar n_\mathrm{min}^\mathrm{fb} \rightarrow 0$), as demonstrated in \cite{Rossi2018} for a MIM system and~\cite{Magrini2021} for a levitated nanoparticle.

\subsection{Imprecision noise analysis}
\label{ssec:excessNoise}

Key challenges when moving from cryogenic to room temperature operation include that not only the quantum cooperativity $C_\mathrm{q}\propto 1/ \bar n_\mathrm{th}$ but also the classical imprecision contribution ($\propto \bar S_{xx}^\mathrm{imp,cl}\cdot \bar n_\mathrm{th}$) in eq.~(\ref{eq:nMinFbAdvanced}) need to be compatible with a $\sim 100$-fold increase of $\bar n_\mathrm{th}$, given that most cryogenic experiments operate at $\sim 4$~K or even lower.
%
%
Several contributions must be considered here.

The first is laser frequency (or phase) noise.
The spectral density of laser frequency fluctuations  $\bar S_{\omega\omega}$ close to $\Omega_\mmr$
can be rescaled to effective displacement using the frequency pull parameter $\dd\omega_\mathrm{c}/\dd x= g_0/x_\mathrm{zpf}$ (cf.~eq.~(\ref{eq:g0})).
If the corresponding imprecision $\bar S_{xx}^\mathrm{imp,ln}=\bar S_{\omega\omega}/(\dd\omega_\mathrm{c}/\dd x)^2 $ dominates over the quantum imprecision, this leads to a minimum occupation on the order of $(\bar S_{\omega\omega} \cdot \bar n_\mathrm{th} \Gamma_\mmr/g_0^2)^{1/2}$ for $C_\mathrm{q}\approx 1$.
Interestingly, a similar limit has been derived and observed for laser sideband cooling  \cite{Schliesser2008, Rabl2009, Safavi2013}.
In our experiment, the frequency noise of the employed laser is specified as $\bar S_{\omega\omega}\lesssim (2 \pi \cdot 1 ~\mathrm{Hz})^2/\mathrm{Hz}$ \cite{NKT}, leading to a neglible contribution as
$\bar S_{\omega\omega} \cdot \bar n_\mathrm{th} \Gamma_\mmr/2 g_0^2\approx 0{.}03$,
much smaller than the quantum imprecision term $1/4 \eta_\mathrm{det} C_\qmr$ (see below).

To analyze a second potential contribution, we need to note that the transduction of the cavity frequency fluctuations into the optical field is inherently nonlinear.
As shown in~\cite{Fedorov2020}, these nonlinearly transduced frequency fluctuations result in a broadband ``thermal intermodulation noise''~(TIN) that can dominate the noise in photodetection
when the linearly transduced fluctuations are small, e.g.~when detecting the intensity of a resonant beam (as in~\cite{Fedorov2020}). 
Conversely, for a homodyne measurement of a resonant beam's phase, employed in our work, the linear contribution is maximal.
Moreover, the leading nonlinear term is not quadratic, but cubic in small fluctuations, which leads to a more favorable scaling of different contributions to TIN (see Supplement~1 for more details).
Finally, because our MIM cavity is deep in the unresolved-sideband regime, the cooling beam provides sideband cooling of a wide range of mechanical modes, which should alleviate the effect of TIN even further \cite{RossiThesis2020}, such that
we expect not to be limited by it.
While TIN-related intensity noise can still be present in the system, it neither appears in the cavity scans (Fig.~\ref{fig:fig1}d)), nor does it manifest as excess heating, as discussed before.

A third relevant source of classical imprecision noise are thermal fluctuations in the cavity mirrors.
This phenomenon has been studied extremely carefully in the context of the large mirrors employed in gravitational wave observatories, and a number of possible mechanisms have been identified \cite{Braginsky2012}.
More compact cavities with millimeter-scale mirrors are typically dominated by the thermal fluctuations of the mirror substrate, displacing the boundary off which the light is reflected.
Since the mirror substrates in general support a discrete set of mechanical resonance modes, the spectrum of these fluctuations is expected to consist of a series of relatively sharp peaks, depending on the $Q$-factor of the mirror modes.
Such spectra have been systematically studied for some two decades \cite{Briant2003,Wilson2012, Nielsen2016}, and indeed even used as mechanical modes in early optomechanics experiments \cite{Briant2003}.
They have also constituted  a limiting factor of MIM experiments, where they can mask quantum effects in the spectral regions where the peaks appear \cite{Purdy2012,Nielsen2017,Mason2019}.
At room temperature, peak displacement spectral densities at the level of $(1-10~\mathrm{am})^2/\mathrm{Hz}$ have been measured for commonly employed mirror geometries \cite{Briant2003,Wilson2012, Nielsen2016}.

\begin{figure}
    \centering
    \includegraphics[width=0.75\textwidth]{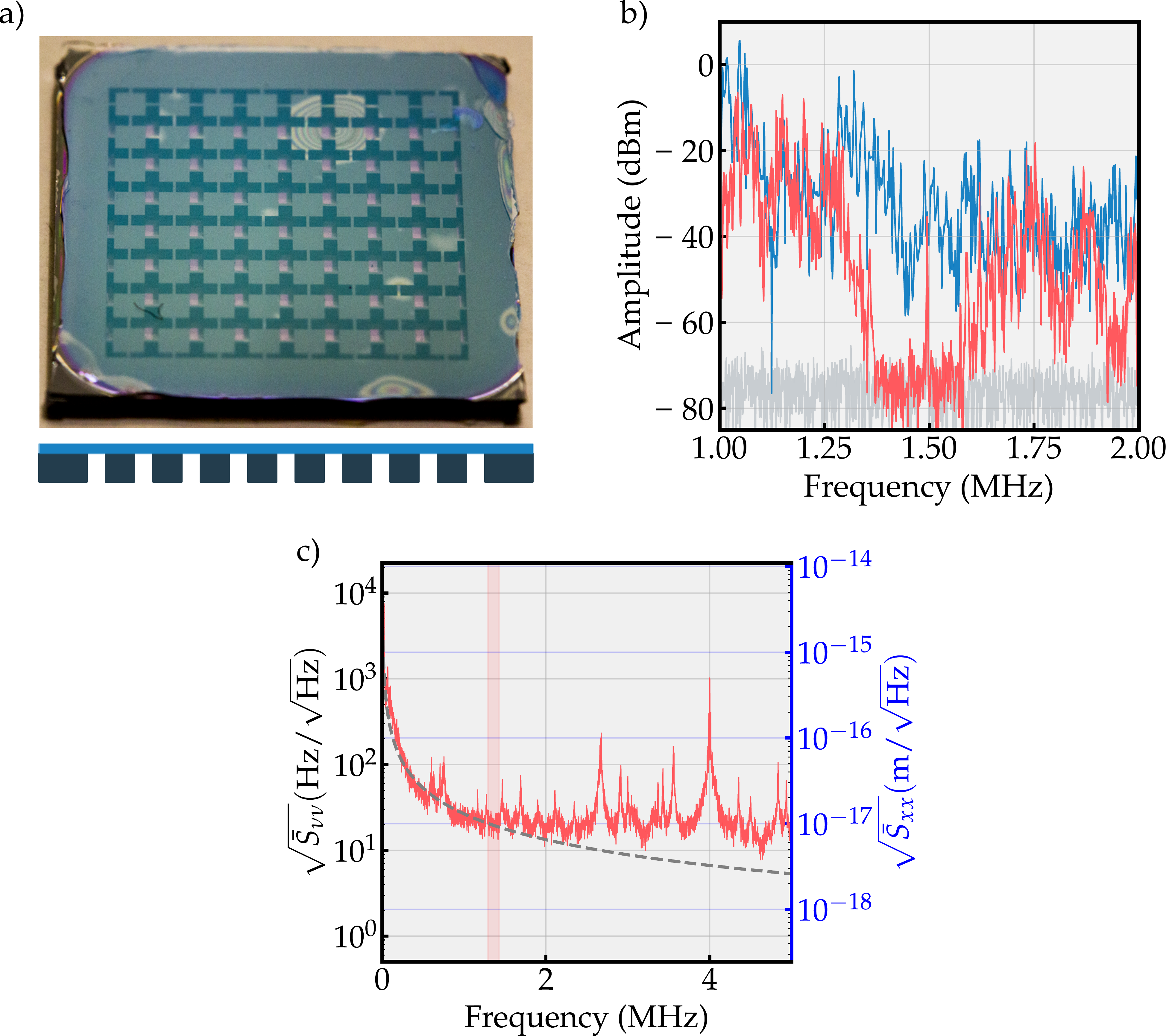}
    \caption{a)~Photograph of an exoskeleton mirror, showing the silicon phononic crystal structure underneath the pyrex wafer, on which a mirror was coated. Interference fringes indicate areas where the pyrex-silicon bonding was not successful. Below is a schematic cross section of the device.
    b)~Spectrum of the driven response on different parts of the exoskeleton mirror: the center (red) exhibits a suppression compared to the frame (blue) for frequencies in a window around  \SI{1.5}{\mega\hertz}. The sharp peak within this window is an externally applied calibration tone. The light-gray trace corresponds to detection noise. c)~Measured homodyne spectrum of the empty optical cavity, presented as frequency noise (left axis, black) and displacement noise (right axis, blue). The dashed gray line indicates (frequency)$^{-1}$ scaling. Both this behaviour at low frequencies and the discernible peaks have been observed in fiber cavities. The light-red shading indicates the frequency region of the exoskeleton bandgap.}
    \label{fig:excessNoise}
\end{figure}

The purpose of our flat mirror's  ``exoskeleton'' (shown in Fig.~\ref{fig:excessNoise}a)) is to reduce the impact of this noise.
Indeed, this silicon
substrate
is structured with a phononic crystal similar to the one employed as membrane support in earlier work \cite{Tsaturyan2013,Yu2014}.
It is modified in such a way that if bonded to the pyrex wafer that carries the mirror reflective coating, the joint structure exhibits an acoustic bandgap in the MHz frequency regime.
We have verified the existence of the bandgap, by measuring the out-of-plane motion of the mirror surface in response to a swept  excitation with a piezoelectric transducer (Fig.~\ref{fig:excessNoise}b)).
Within the bandgap, we therefore expect no substrate thermal noise peaks to appear at a relevant level.

Informed by earlier studies of guided acoustic-wave Brillouin scattering in single-mode fiber \cite{Shelby1985}, we would expect the acoustic resonances of the fiber mirror to exhibit a sparse spectrum and in particular remain confined to high frequencies $\gtrsim 10~\mathrm{MHz}$.
A dedicated study \cite{Brachmann2016} of a fiber cavity has revealed however, that e.g.~fiber bending modes can occur already in the MHz-regime, and more importantly, also the existence of a broadband noise at the level of about $(20~\mathrm{am}) ^2/\mathrm{Hz}$ in the MHz frequency region.

We have independently analysed the backgound noise in our composite exoskeleton/fiber-mirror cavity, with a dedicated measurement of its apparent displacement noise with the membrane removed.
The result is shown in Fig.~\ref{fig:excessNoise}c).
Not only does it display peaks already above 2~MHz, it does indeed also show a broadband noise, which reaches the level of $\sim(10~\mathrm{am}) ^2/\mathrm{Hz}$ at $1~\mathrm{MHz}$, well compatible with earlier measurements  \cite{Brachmann2016} (note that we only use one fiber mirror).
We interpret the absence of any signature of the exoskeleton's bandgap in this joint noise spectrum to be a consequence of the fact that the latter is dominated by broadband noise of the fiber mirror.
Whereas this noise is certainly sub-ideal, we can tolerate the associated level of classical imprecision (with $\bar S_{xx}^\mathrm{imp,cl}\cdot \bar n_\mathrm{th} \Gamma_\mmr/2 x^2_\mathrm{zpf}\approx 0{.}5)$) in the current experiments,
since the quantum noise contribution is much larger (see below).

Quantum noise in the detection is the last contribution to consider.
For a resonant probe in the unresolved sideband regime, the equivalent displacement noise power spectral density is given by
$\bar S_{xx}^\mathrm{imp,q} =\hbar^2/ \eta_\mathrm{det} C_\mathrm{q} \bar S_{FF}^\mathrm{th}=x_\mathrm{zpf}^2/2\eta_\mathrm{det}C_\mathrm{q} \gamma$.
Here, the overall detection efficiency is determined by the product $\eta_\mathrm{det}=\varepsilon\cdot \eta_\mathrm{c} \cdot |\beta| \cdot \mathcal{V} \cdot P_\mathrm{QE}$, with the efficiency $\varepsilon\cdot\eta_\mathrm{c} $ for coupling cavity photons into the guided fiber mode, fiber losses $|\beta|$, homodyne fringe visibility $\mathcal{V}$, and the photodetectors' quantum efficiency $P_\mathrm{QE}$.
We achieve
$\eta_\mathrm{det} \approx 1{.}2\%$, dominated by a relatively low cavity mode-matching $\varepsilon\approx 4\%$,
see Supplement~1 for details.
We determine the quantum cooperativity $C_\mathrm{q}=4g^2/\kappa \gamma$ of the probe from the measured power ratio of the two beams, their known detunings, and the cooling beam's coupling rate $g_\mathrm{c}$ as determined by the dynamical backaction measurements described in section~\ref{s:approach}.

Our laser does not provide enough optical power to reach unity cooperativity with the probe, and simultaneously feed the cooling beam and local oscillator with sufficient power. 
We therefore operate with a relatively modest probe power, corresponding to $C_\mathrm{q}=0{.}1$.
We then have enough optical power available for the local oscillator to overwhelm the detectors’ electronic noise with shot noise. 
Whereas higher quantum cooperativity would certainly be desirable, the achieved value constitutes a significant advance over
many tabletop
room-temperature experiments, reaching $C_\mathrm{q}\approx 10^{-5}$ \cite{Purdy2017}, $C_\mathrm{q}<10^{-4}$ \cite{Rochau2021}, $C_\mathrm{q}\approx 4 \cdot 10^{-3}$\cite{Guo2019} and $C_\mathrm{q}\approx 5 \cdot 10^{-3}$\cite{Sudhir2017},
and is
on par
with that in \cite{Cripe2019}.
With this available probe power, the expected quantum imprecision of
$\bar S_{xx}^\mathrm{imp,q} \approx (210\,\mathrm{am})^2/\mathrm{Hz}$
is still well above the mirror noise.

\subsection{Cooling results}
To cool the membrane mode close to
its motional
ground state, we employ the setup shown in Figure \ref{fig:Fig_2}a), and described already in section \ref{s:approach}.
Additional details can be found in Supplement~1.
%
Cooling proceeds via the combined effect of laser sideband and feedback cooling.
That is, the cooling beam is detuned by $\Delta_\mathrm{c}/2\pi\approx -80\,\mathrm{MHz}$ from the cavity resonance, and its dynamical backaction
damps
the mechanical motion.
The corresponding ``anchor'' spectrum of mechanical fluctuations is recorded by homodyne detection of the second, probe beam.
The broadened mechanical resonance width $\Gamma_\mathrm{m}^\mathrm{tot}/2\pi\approx$ \SI{52}{\hertz} can be straightforwardly extracted from the homodyne spectrum via a Lorentzian fit.
Together with independently determined optical ($\Delta_\mathrm{c}$, $\kappa$) and mechanical  ($\Omega_\mathrm{m}$, $\Gamma_\mathrm{m}$) parameters, we can
estimate
the resulting occupation $\bar n$ of the mode after sideband cooling from equations (\ref{e:barn}) and (\ref{e:barnmin}).
%
Here, the bare mechanical damping $\Gamma_\mathrm{m}$ is obtained from ringdown measurements in a different setup and corrected for the slightly different pressure, see
Supplement~1
for details.
The calculated sideband-cooled mode occupation $\bar n \approx 1070$ quanta is then used to calibrate the  homodyne spectrum, which is originally recorded in electrical (voltage) units.
That is, the measured homodyne spectrum $
    \bar S_{yy}(\Omega) =  \bar S_{xx}(\Omega) +   \bar S_{xx}^\impmr $
is calibrated in displacement units so that 
\begin{align}
   \label{eq:barn}
   \bar n = \int_{0}^\infty \frac{\bar S_{xx}(\Omega)}{2 x_\zpf^2} \frac{\dd\Omega}{2 \pi} - \frac{1}{2}
\end{align}
holds, whereby $\bar S_{xx}(\Omega)=|\chi_\mmr^\tot(\Omega)|^{2}  \bar S_{FF}^\tot$ and $\chi_\mmr^\tot(\Omega)$ is the Lorentzian mechanical susceptibility modified by
dynamical backaction.
%

We now turn on feedback cooling of the membrane mode.
To this end, the homodyne signal $y(t)$ is filtered and amplified, and subsequently used to modulate the radiation-pressure force  $F_\fbmr(t) = h_\fbmr(t) \ast y(t)$ exerted by the cooling beam via the acousto-optic modulator (Figure \ref{fig:Fig_2}a)). 
In the frequency domain, the filter's transfer function is of the form $h_\fbmr(\Omega) = h_\mathrm{main}(\Omega) + h_\auxmr(\Omega)$, with
\begin{align}
    h_\mathrm{main}(\Omega) = G_\fbmr e^{i ( \Omega \tau_\fbmr - \phi_\fbmr)}\, \left[ \frac{\Gamma_\fbmr \Omega}{\Omega_\fbmr^2 - \Omega^2 - i\Gamma_\fbmr \Omega} \right]^2 \, .
    \label{eq:1}
\end{align}
Here $G_\fbmr$ is the overall feedback gain, the main filter central frequency is $\Omega_\fbmr/2\pi = \SI{1.34}{\mega\hertz}$ and the bandwidth $\Gamma_\fbmr/2\pi = \SI{77.86}{\kilo\hertz}$.
The total delay is measured to be $\tau_\fbmr \approx \SI{300}{\nano\second}$, and the phase $\phi_\fbmr$ is adjusted so as to give $\arg(h_\fbmr(\Omega_\mmr^\tot)) \approx \pi/2$ (where $\Omega_\mmr^\tot = \Omega_\mmr + \delta\Omega_\mmr^\mathrm{p} + \delta\Omega_\mmr^\mathrm{c}$ and $\delta\Omega_\mmr^\mathrm{p}$ is the optical spring from the probe),
which corresponds to
$F_\fbmr$ approximating
a friction force on the membrane. 
The function $h_\auxmr(\Omega)$ represents a narrow auxiliary filter designed to stabilize a mechanical mode just outside of the bandgap, excited by the main filter, while maintaining $h_\fbmr(\Omega_\mmr^\tot) \approx h_\mathrm{main}(\Omega_\mmr^\tot)$. More information on the filter and its implementation can be found in
Supplement~1.

Closing the feedback loop leads to a modified susceptibility of the mechanical mode to forces,
\begin{align}
    \chi_\fbmr(\Omega)=\frac{\chi_\mmr^\tot(\Omega)}{1-\chi_\mmr^\tot(\Omega) h_\fbmr(\Omega)}.
\end{align}
%
Furthermore, also the measurement (imprecision) noise is now fed back as a force, so that 
\begin{align}
    \bar S_{xx}(\Omega) = |\chi_\fbmr(\Omega)|^2 \left( \bar S_{FF}^\tot + |h_\fbmr(\Omega)|^2 \bar S_{xx}^\impmr \right),
    \label{eq:Sxxfb}
\end{align}
whereas the measured (in-loop) homodyne spectra are modified as 
\begin{align}
    \bar S_{yy}(\Omega) =  |\chi_\fbmr(\Omega)|^2 \left( \bar S_{FF}^\tot + |\chi_\mmr^\tot(\Omega)|^{-2}  \bar S_{xx}^\impmr \right).
    \label{eq:Syyfb}
\end{align}

\begin{figure}
     \centering
     \includegraphics[width=0.9\textwidth]{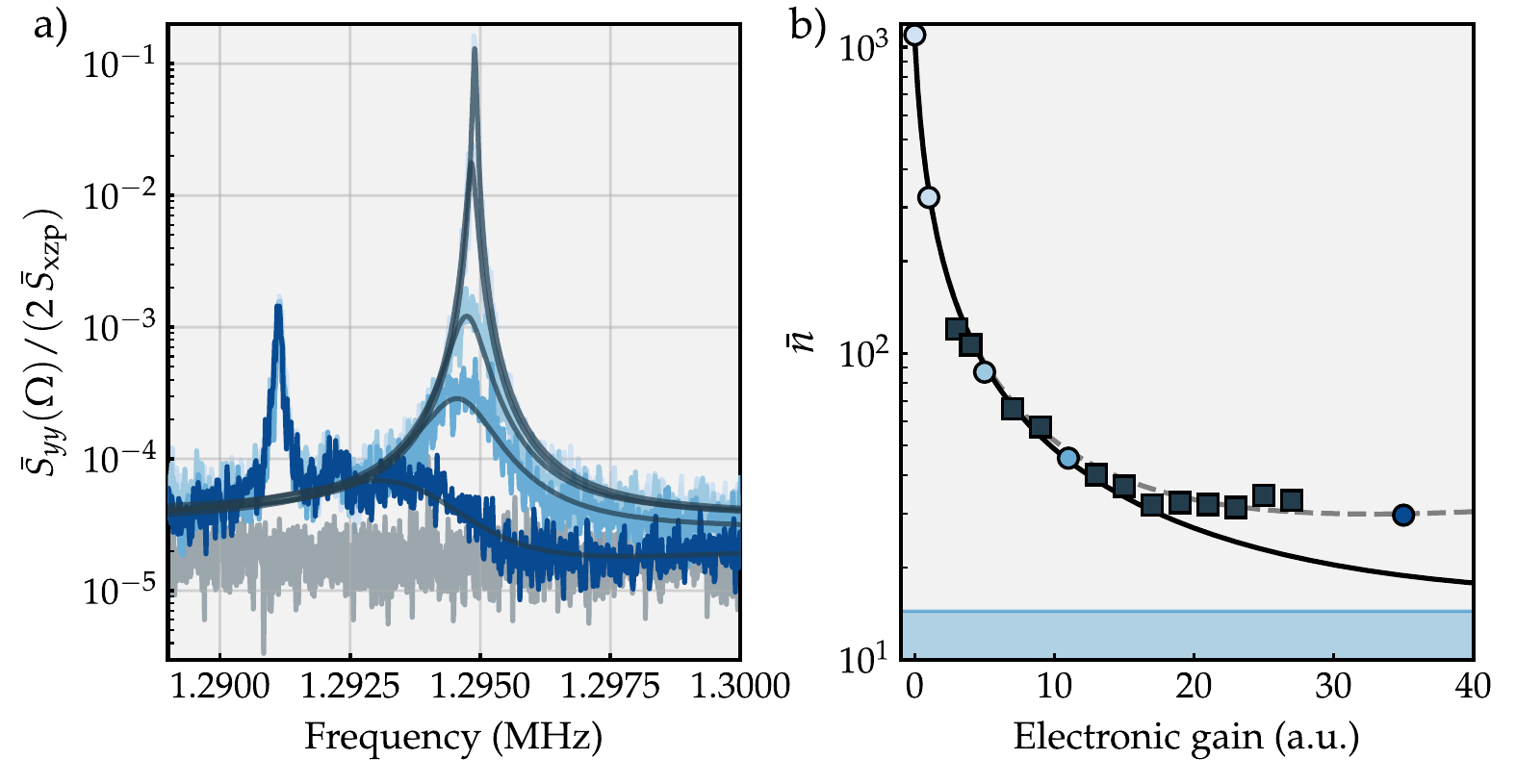}
     \caption{a)~Measured mechanical spectra for progressively higher feedback gains from light- to dark-blue.
     The lightest trace is obtained with pure sideband cooling and is used as a calibration anchor.
     The right peak is the mode of interest, corresponding to the higher-frequency defect mode in Fig.~\ref{fig:fig1}c), now shifted by the optical spring.
     The gray trace corresponds to shot noise of the light. Dark-gray lines are fits to the data. The feature to the left is a stiff mechanical mode not susceptible to radiation pressure effects, and is excluded from the fit. b) The occupation inferred from fits to the
     sideband- and
     feedback-cooled spectra as a function of electronic gain. Circles are color-coded to the corresponding traces in a), whereas squares correspond to data not shown in a) for clarity. The solid black line is the theoretical prediction for quantum-limited imprecision (cf. Supplement~1), whereas the dashed grey line includes excess noise. The top of the blue-shaded region indicates the ideal-filter threshold,
     $\bar n_\mathrm{min}^\mathrm{fb}$.
     }
     \label{fig:fig4}
 \end{figure}

Figure \ref{fig:fig4} shows the homodyne spectra measured with the feedback loop closed, and plotted normalized to the peak spectral density in the ground state
$\bar S_\xzp \equiv 4 x_\zpf^2/\Gamma_\mmr$. 
Here we use the same conversion factor for the homodyne spectra from electrical (voltage) to displacement units as obtained for the only sideband-cooled, anchor spectrum described above, under the reasonable assumption that this conversion is not affected by changes in the feedback gain.

We fit the model of eq.~(\ref{eq:Syyfb}) to these spectra, adjusting the free parameters $G_\fbmr$, $\phi_\fbmr$ and the measurement imprecision in units of quanta,
$\bar n_\impmr = \bar S_{xx}^\impmr / (2 \bar S_\xzp)$.
The total force noise in units of quanta,
$\bar n_\tot = \bar S_{FF}^\tot \bar S_\xzp/(8\hbar^2)$,
is fixed to the value expected from the sum of thermal bath and quantum backaction heating.
Other parameters of the model are extracted from the homodyne spectrum in the absence of feedback ($\Omega_\mathrm{m}^\tot$, $\Gamma_\mathrm{m}^\tot$) or measured independently. The best-fit values of the fitted parameters as a function of gain are shown in Supplement~1. 
We plug
these
best-fit
values into eq.~(\ref{eq:Sxxfb}), and integrate the resulting displacement spectral density, as in eq.~(\ref{eq:barn}), to obtain the mechanical occupation $\bar n$ equivalent to the measured spectra.
The results are shown in Fig.~\ref{fig:fig4}b) as a function of feedback gain. As a sanity check, the described analysis is done also for the anchor spectrum, constraining $G_\fbmr \approx 0$, and the estimated sideband-cooled occupancy is obtained to within $\approx \SI{1}{\percent}$.

%
%
The lowest occupancy we have obtained is $\bar n = 30 \pm 8$, where the confidence interval is derived by propagating the {statistical} uncertainty in the area of the anchor spectrum, which is used for calibrating the feedback-cooled spectra.
The uncertainty propagated from those of the fit parameters for the feedback-cooled spectrum is negligible in comparison.
We also expect systematic effects such as power or polarization drifts to not contribute significant additional uncertainty.
From fits to the calibrated spectra, we extract a very low number of imprecision noise quanta, with the average over all gain values being $\bar n_\impmr = {(3.2\pm 0{.}6)\times 10^{-5}}$, corresponding to $\bar S_{xx}^\impmr ={(370\pm 20\,\mathrm{am})^2/\mathrm{Hz}}$. Still, this number is above the expected quantum imprecision noise. 
%
%
Correspondingly, for higher gains, the achieved occupation reaches the limit of 30 quanta expected from integrating eq.~(\ref{eq:Sxxfb}) in the presence of excess imprecision $\bar S_{xx}^\mathrm{imp,cl}$, as compared to $\bar n_\mathrm{min}^\mathrm{fb} \approx 15$ were $\bar S_{xx}^\mathrm{imp,cl}$ negligible (see Fig.~\ref{fig:fig4}b)). Two possible causes of the excess noise include cooling beam bleedthrough to the probe homodyne spectrum, due to imperfect polarization control, and residual TIN. However, other sources cannot be ruled out based on an extended analysis discussed in Supplement~1.

%
%
%
%



\section{Discussion}

%
Regarding improvements to the experiment's performance,
first and foremost, better alignment should help remedy the most significant optical losses due to poor mode matching between the cavity mode and the single-mode fiber.
{This should be possible by using a more precise alignment stage that positions the fiber with the fiber mirror before gluing it in place and, in particular, sets the pitch and yaw angles relative to the exoskeleton mirror.
Namely, while the maximally attainable mode-matching efficiency is determined by the decentration of the fiber mirror vertex with respect to the fiber core,}
efficiencies on the order of $\SI{80}{\percent}$ have been achieved with fiber cavities \cite{Hunger2010}, a $\sim20$-fold improvement over the current performance.
Improved mode matching
would simultaneously allow higher intracavity powers, even with the rest of the experiment unchanged, so that $C_\mathrm{q}\sim 2$ can be reached and combined with homodyne detection.
{It would also mitigate two of the potential contributions to classical imprecision, which stem from imperfect control of the cooling beam polarization (by allowing for a larger LO power) and from TIN (by enabling stronger sideband cooling).}
%
In the absence of other imprecision noise, realistic further improvements to an overall detection efficiency of $\eta_\mathrm{det}=\SI{50}{\percent}$
then enable $\bar n<1$ even in the presence of fiber mirror noise, according to eq.~(\ref{eq:nMinFbAdvanced}).
To reach $\bar n<0.5$, the mirror noise could be suppressed by a factor of 10. This might be accomplished by replacing the fiber mirror with a second exoskeleton mirror,
featuring a concave indentation made by laser ablation~\cite{Kuhn2014} or by solvent vapor reflow and reactive ion etching~\cite{Jin2022_arxiv}.

In comparison with other experiments performing feedback cooling of room-temperature devices, our achieved occupancy is orders of magnitude below what has been demonstrated with most suspended mechanical systems \cite{Kleckner2006, Schafermeier2016, Pluchar2020} and on par with state-of-the-art nanodevices ($\bar n=27$) \cite{Guo2019}, as well as the test masses in advanced LIGO ($\bar n=11$) \cite{LIGO2021}.
Lower occupancies ($\bar n<1$) have recently been achieved with nanoparticles held in optical traps \cite{Delic2020, Magrini2021}, which are  some five orders of magnitude lighter than the resonators studied here.
We see a great potential of the membrane platform for room temperature quantum experiments, due to its relative simplicity and great versatility. 
In addition to the many possibilities arising from a quantum-enabled optomechanical interface (see introduction), membrane resonators can  be functionalized to couple to charge \cite{Bagci2014, Andrews2014} or spin \cite{Fischer2019,Kosata2020}, and can be decorated with samples such as viruses or metal nanoparticles \cite{Halg2021}.
This opens many application prospects which may benefit from initializing the mechanical system in a low-entropy state, e.g. by allowing to discern small displacements in phase space. 
A particularly interesting prospect would also be to implement a compact, wavelength-agnostic, room-temperature source of broadband ponderomotive squeezing~\cite{Aggarwal2020}.

\section*{Funding}
 This work was supported by the European Research Council projects Q-CEOM (grant no. 638765) and PHOQS (grant no. 101002179), the Danish National Research Foundation (Center of Excellence “Hy-Q”), the Independent Research Fund Denmark (grant no. 1026-00345B), the Novo Nordisk Foundation (grant no. NNF20OC0061866) as well as the Swiss National Science Foundation (grant no. 177198). The project has furthermore received funding from the European Union’s Horizon 2020 research and innovation program under grant agreement No. 722923 (Marie Curie ETN - OMT).

\section*{Acknowledgements}
The authors would like to thank David Hunger of Karlsruhe Institut f\"{u}r Technologie for access to and training on their setup for fabricating and characterizing fiber mirrors. The authors also thank Eva Weig of Technische Universit\"{a}t M\"{u}nchen for equipment and help in preparing the fibers and Irene S\'{a}nchez Arribas for assisting in fiber fabrication. Finally, the authors acknowledge Rik van Herk for acquiring and analyzing the data for the mechanical quality factor as a function of pressure, shown in Supplement 1.

\section*{Disclosures}
The authors declare no conflicts of interest.

\section*{Data availability}
Data underlying the results presented in this paper are not publicly available at this time, but may be obtained from the authors upon reasonable request.



\bibliography{RT_fiber_refs}

\end{document}


\maketitle

\section{Feedback cooling theory and analysis}
 
The effect of the feedback, including the noise, can be understood by evaluating the (symmetrized) spectral density of mechanical position fluctuations, which can be written in the frequency domain as
\begin{equation}
    \bar S_{xx}(\Omega) = |\chi_\fbmr(\Omega)|^2 \left( \bar S_{FF}^\tot + |h_\fbmr(\Omega)|^2 \bar S_{xx}^\impmr \right) 
    \label{eq:Sxx}
\end{equation}
when the feedback loop is closed.
Here, $\chi_\fbmr(\Omega)$ is the mechanical susceptibility modified by the feedback, 
$\bar S_{FF}^\tot$ is the spectral density of force fluctuations driving the mechanics, and $\bar S_{xx}^\impmr$ is the spectral density of $x_\impmr(t)$.
In a toy model known as cold damping \cite{Cohadon1999}, the feedback filter is approximated as a differentiator,
\begin{align}
    h_\fbmr(\Omega)\approx i m \Omega \Gamma_\mmr g_\mathrm{fb} \, ,
\end{align}
in the relevant frequency range, where $g_\mathrm{fb}$ is a dimensionless gain parameter.
The resultant velocity-dependent feedback force leads to increased damping in the effective mechanical susceptibility
\begin{align}
    \chi_\fbmr(\Omega)=\frac{1}{m}\frac{1}{\Omega_\mmr^2-\Omega^2-i \Gamma_\mmr \Omega (1+g_\mathrm{fb})} \, .
\end{align}
Sufficient gain $g_\mathrm{fb}\gg 1$ can then reduce the residual mechanical occupation as obtained from the mechanical position variance under the assumption of equipartition,
\begin{align}
       \bar n = \int_{0}^\infty \frac{\bar S_{xx}(\Omega)}{2 x_\zpf^2} \frac{\dd\Omega}{2 \pi} - \frac{1}{2} \, .
       \label{eq:si_n1}
\end{align}
%
%
An instructive discussion of this topic in the (complementary) framework of a stochastic master equation is also given in \cite{Bowen2016}. 
The expressions agree in the here applying limit of large (classical) cooperativity.

Here, instead, we provide a quick derivation of eq.~(7) of the main text using the cold damping model. We have already assumed that $\bar S_{FF}^\tot$ and $\bar S_{xx}^\impmr$ are flat in the range where $|\chi_\fbmr(\Omega)|^2$ is significant. The contribution to the occupancy from the total force fluctuations features the integral
\begin{align}
    \int_0^\infty |\chi_\fbmr(\Omega)|^2 \dd \Omega = \frac{1}{m^2} \int_{-\infty}^\infty \frac{\dd \Omega}{(\Omega_\mmr^2 - \Omega^2)^2 + \Gamma_\mmr^2(1 + g_\fbmr)^2 \Omega^2} \, ,
\end{align}
whereas the integral that figures in the contribution from the measurement imprecision is
\begin{align}
    \int_0^\infty |\chi_\fbmr(\Omega)|^2 \Omega^2 \dd \Omega = \frac{1}{m^2} \int_{-\infty}^\infty \frac{\Omega^2 \dd \Omega}{(\Omega_\mmr^2 - \Omega^2)^2 + \Gamma_\mmr^2 (1 + g_\fbmr)^2 \Omega^2} \, .
\end{align}
With the change $\Omega \rightarrow z, z \in \mathbb{C}$, we note that both
integrands are analytical in the entire upper half of the complex $z$-plane and converge
to 0 in the upper half-plane, including the real axis, for $|z| \rightarrow \infty$. Thus the integrals along the half-circle in the upper half-plane vanish (Jordan’s lemma) and the integrals along the real axis can be calculated using the residue theorem. See Ref.~\cite{Schliesser2009} for the full calculation. Since both of the integrands are also even, the values of the integrals are simply half of those taken over the entire real axis, and the results are given by
\begin{align}
    \int_0^\infty |\chi_\fbmr(\Omega)|^2 \dd \Omega
    & = \frac{\pi}{2 \, m^2 \, \Omega_\mmr^2 \, \Gamma_\mmr (1 + g_\fbmr)} \, ,
    \label{eq:si_integral1}
\end{align}
and
\begin{align}
    \int_0^\infty |\chi_\fbmr(\Omega)|^2 \Omega^2 \dd \Omega =
    \frac{\pi}{2 \, m^2 \, \Gamma_\mmr (1 + g_\fbmr)} \, .
    \label{eq:si_integral2}
\end{align}
Combining these
with the appropriate prefactors according to eqs.~(\ref{eq:Sxx})--(\ref{eq:si_n1}) yields for the phonon occupancy
\begin{align}
    \bar n = \frac{1}{8 x_\zpf^2} \left( \frac{1}{m^2 \, \Omega_\mmr^2 \, \Gamma_\mmr} \cdot \frac{\bar S_{FF}^\tot}{1 + g_\fbmr} + \Gamma_\mmr \cdot \frac{g_\fbmr^2 \, \bar S_{xx}^\impmr}{1 + g_\fbmr} \right) - \frac{1}{2}\, .
\end{align}
In the limit $g_\fbmr \gg 1$, this expression reaches a minimum for
\begin{align}
  g_\fbmr = \frac{1}{m \, \Omega_\mmr \, \Gamma_\mmr} \sqrt{\frac{\bar S_{FF}^\tot}{\bar S_{xx}^\impmr}} \, ,
\end{align}
resulting in $\bar n_\mathrm{min}^\mathrm{fb}$ given by eq.~(7).

\subsection{Implemented controller}

\begin{figure}[h]
    \centering
    \includegraphics[width=0.75\textwidth]{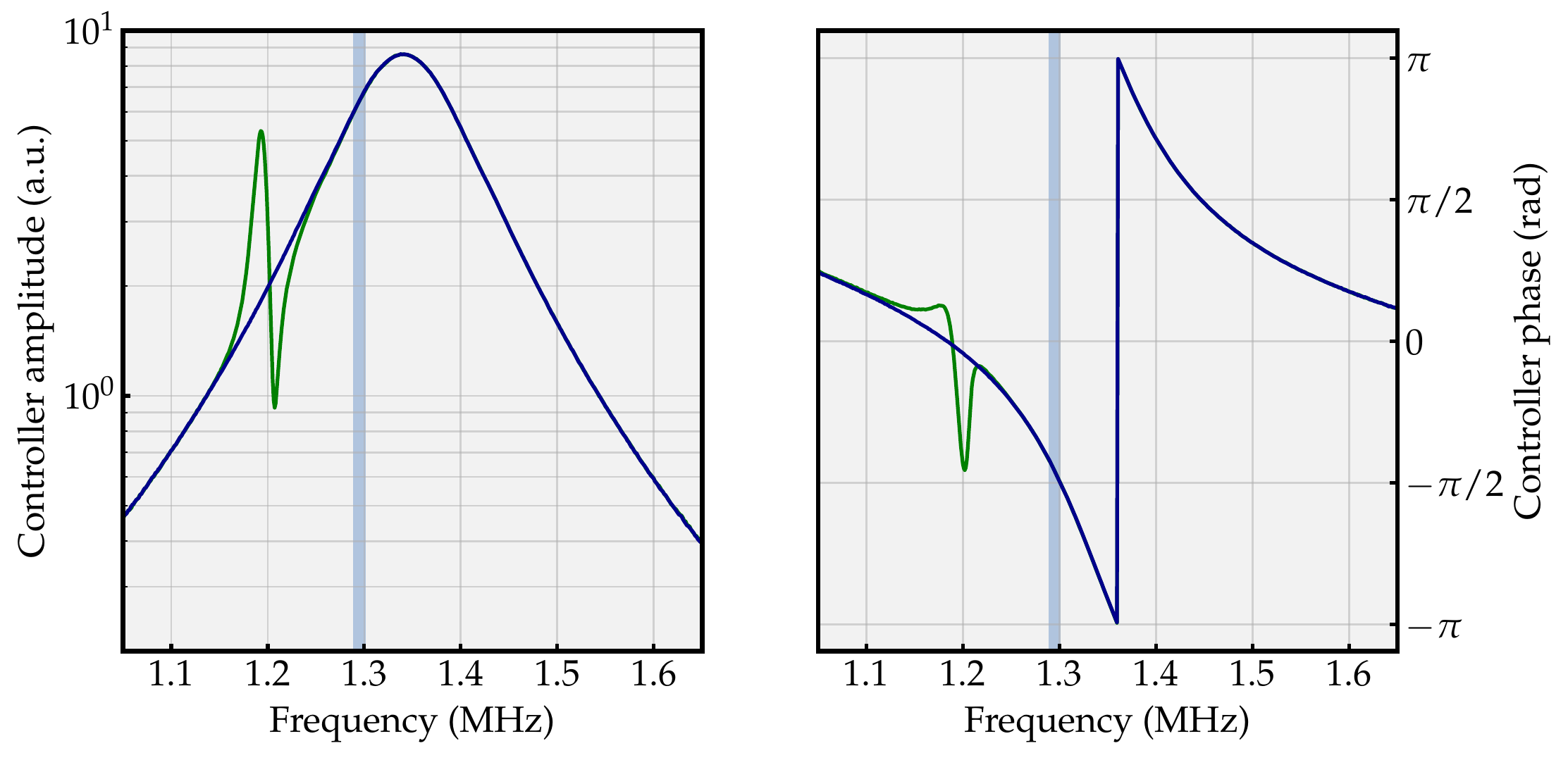}
    \caption{The amplitude and phase of the Red Pitaya controller. The blue curve corresponds to just the main filter, while the filter used for the feedback cooling run discussed in the main text (green) contains also the auxiliary filter used to stabilize an out-of-bandgap mode at $\Omega/2\pi \approx \SI{1.195}{\mega\hertz}$. The shaded region tentatively indicates the mode of interest, shifted by dynamical backaction.
    }
    \label{fig:si_RP_function}
\end{figure}

Our experimental controller is inspired by that in Ref.~\cite{Rossi2018}. It builds on the optimal controller for a single mode, while accounting for finite loop delay and the fact that the system is actually multimode. Specifically, the total feedback transfer function from the output of the cavity to applying the feedback force, including the optical path, detection electronics, electronic controllers and cables and optical amplitude modulation, can be written as
\begin{equation}
    h_\fbmr(\Omega) = h_\mathrm{main}(\Omega) + h_\auxmr(\Omega) = G_\fbmr e^{i ( \Omega \tau_\fbmr - \phi_\fbmr)} \left[ \frac{\Gamma_\fbmr \Omega}{\Omega_\fbmr^2 - \Omega^2 - i\Gamma_\fbmr \Omega} \right]^2  + h_\auxmr(\Omega)\, . \nonumber
\end{equation}
Here $\tau_\fbmr$ is the loop delay time and $G_\fbmr$ and $\phi_\fbmr$ are the feedback gain and phase offset respectively, both tuned within the controller (the Red Pitaya IQ module). The phase $\phi_\fbmr$ is adjusted to yield  $\arg(h_\fbmr(\Omega_\mmr^\tot)) \approx \pi/2$, i.e. such that the feedback exerts a friction force on the membrane.
The main filter center frequency is $\Omega_\fbmr/2\pi = \SI{1.34}{\mega\hertz}$ and the bandwidth is $\Gamma_\fbmr/2\pi = \SI{77.86}{\kilo\hertz}$. The eventual need for auxiliary filters stems from the fact that the feedback can be $\pi$ out of phase at the frequencies of other (bandgap or out-of bandgap) mechanical modes, causing instability and limiting the maximum gain. The fourth-order bandpass in principle minimizes the filter response outside of the bandgap, but in practice we observe a mode just below the bandgap, at $\Omega/2\pi \approx \SI{1.195}{\mega\hertz}$ being excited for high enough gains. Fig.~\ref{fig:si_RP_function} shows the measured transfer function of the Red Pitaya for just the main filter (blue) and with an additional narrow filter used to stabilize the mentioned out-of-bandgap mode (green). The effect of the auxiliary filter is negligible around the frequency of the mode of interest
(indicated as a shaded range of frequencies below $\Omega_\mmr/2\pi = \SI{1.3}{\mega\hertz}$ because of sideband cooling).
Note that in this measurement the phase offset is arbitrary. In the actual cooling experiment, the offset is adjusted so as to give the total controller phase of $\pi/2$ at $\Omega = \Omega_\mmr^\tot$, accounting also for the additional loop delay caused by the amplitude modulator, detection electronics and cables and the optical path. In practice, this offset optimization is done by minimizing the resonance frequency shift seen in the mechanical spectra.
%

\subsection{Best-fit values for feedback-cooled spectra}

\begin{figure}[h]
    \centering
    \includegraphics[width=0.97\textwidth]{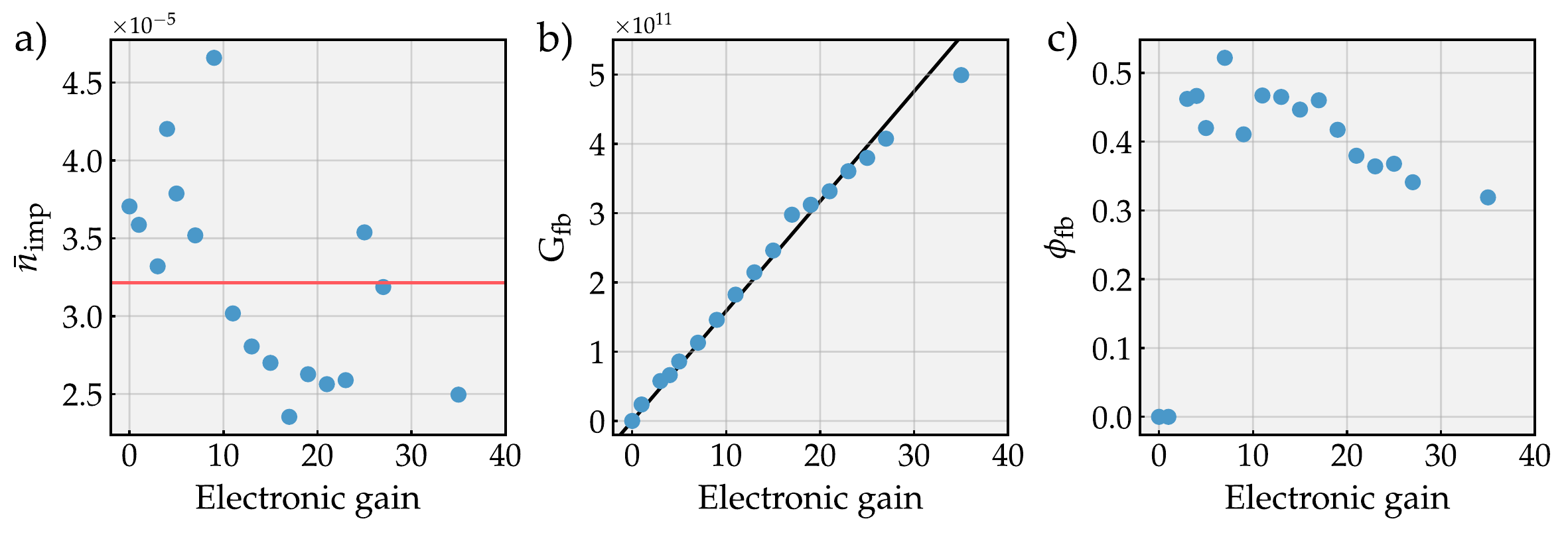}
    \caption{The best-fit values for a)~imprecision noise quanta, b)~feedback gain and c)~feedback phase offset, as a function of the Red Pitaya gain. The gain is fitted to a linear function with zero $y$-intercept (black line), and the slope is used to calibrate the solid and dashed lines in Fig.~4b) of the main text. The dashed line in Fig.~4b) also uses the average of the fitted imprecision values (red line).    
    }
    \label{fig:si_fitted_params}
\end{figure}

As explained in the main text, the spectra in Fig.~4a) are fitted to the model of eq.~(13) with the following fit parameters: the feedback gain and phase offset, $G_\fbmr$ and $\phi_\fbmr$ respectively, and the number of impresicion noise quanta, $\bar n_\impmr$. Conversely, the number of total force noise quanta is fixed to the sum of the thermal bath and quantum backaction heating. The best-fit values for the three fitted parameters obtained for individual spectra are shown in Fig.~\ref{fig:si_fitted_params} as a function of the corresponding gain values in the Red Pitaya IQ module.

The best-fit values for $G_\fbmr$ are then also fitted to a linear function with imposed zero $y$-intercept. Namely, the slope of this fit is used to properly scale the theoretical predictions for the occupancy, both for quantum-noise-limited imprecision and in the case of excess imprecision noise (corresponding to the solid black and dashed grey line in Fig.~4b), respectively). Additionally, the dashed line in Fig.~4b) also uses the average of the fitted imprecision values. 


\section{Fabrication}


\subsection{Membrane}

\begin{figure}[h]
    \centering
    \includegraphics[width=0.95\textwidth]{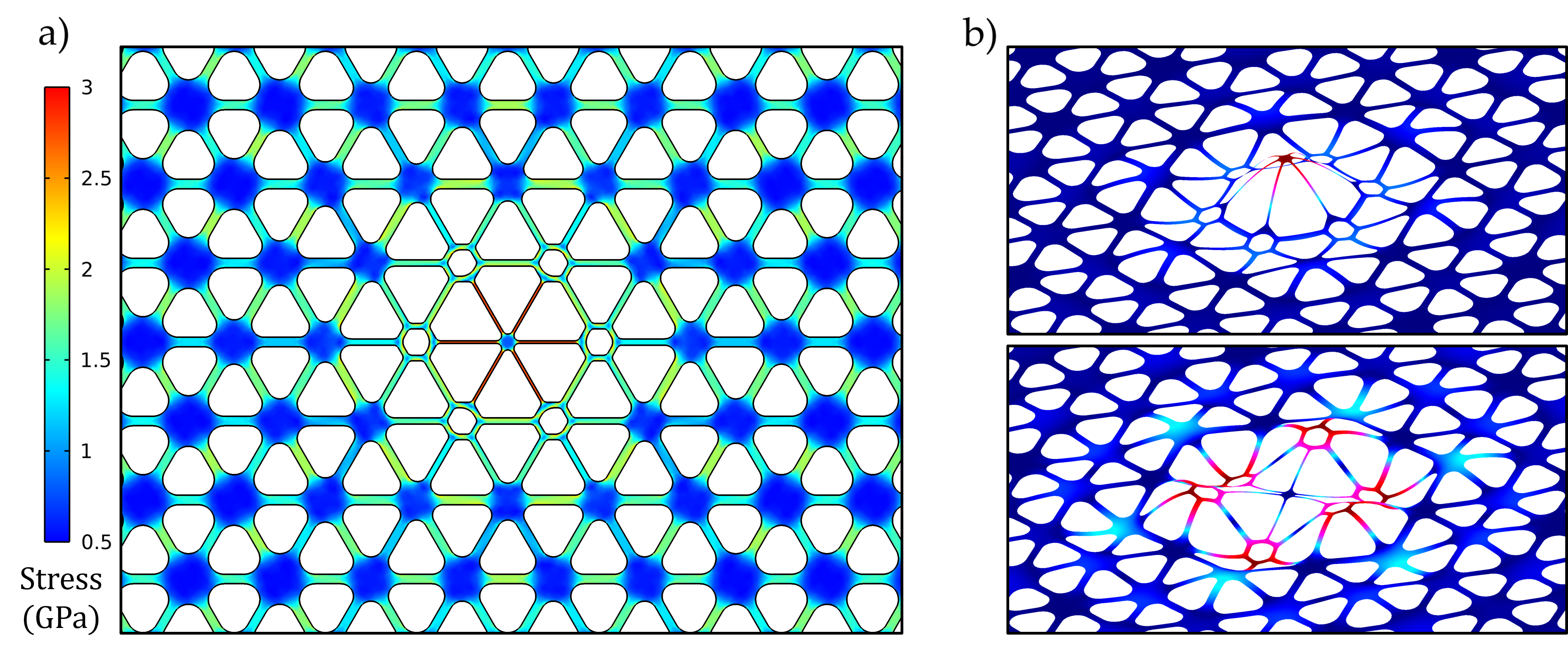}
    \caption{Simulations for the dandelion-class phononic membrane resonator used in this work, performed with COMSOL Multiphysics. 
    (a) The predicted stress redistribution profile following release during the fabrication process. 
    (b) The two defect modes simulated within the out-of-plane bandgap of our phononic crystal. The top mode, with maximum displacement at the defect center, is the mode-of-interest in this work.  
    }
    \label{fig:ddl_sim}
\end{figure}

We utilize a class of phononic crystal membrane resonators first introduced in \cite[Supplementary Information]{Halg2021}, now referred to as \textit{Dandelion}-class resonators. This resonator design is inherently a six-point trampoline embedded in a phononic crystal membrane in a way that realizes soft-clamped mechanical modes. Generally, these membranes possess two mechanical modes found within the out-of-plane phononic bandgap (see Fig.~\ref{fig:ddl_sim}b)). The placement of both modes is controlled through engineering of the defect geometry to modify the mass and stress profile of each mode. The final design chosen for this work was settled on for three primary reasons: (1)~the mechanical mode of interest was centered in a relatively large bandgap, (2)~the central defect was increased in size to be over \SI{30}{\micro\meter} in diameter, and (3)~the predicted stress redistribution of the membrane following release stayed within the functional range of 0.5--\SI{3.0}{\giga\pascal}. Dandelion-class resonators will be explored more thoroughly in a forthcoming manuscript.

The fabrication procedure is nearly identical to the one utilized in~\cite{Tsaturyan2017}. The only significant difference is that lithographic exposure of the patterns was done using a Heidelberg MLA-150 maskless lithography tool, rather than standard hard mask lithography. This was invaluable for the quick fabrication of designs slightly modified between each wafer.

\subsection{Fiber mirror}
Typical fibers have their glass core surrounded by a thin polymer layer and, depending on the configuration, there are usually additional protective layers. The bare fiber is very fragile and thus at least the innermost protection is needed. However, the typical polymer used to protect the fiber will not withstand the temperatures the fiber would be exposed to during the deposition of the high quality mirror coating. Instead, we choose to use copper-coated fibers (IVG fiber CU-1300, \SI{125}{\micro\meter} diameter) that are otherwise identical to the usual ones.

The initial step in the fabrication is removing all protective coatings and cleaving the fiber facet. Ideally, the fiber facet should be perpendicular to the fiber core, such that the cavity axis coincides with it. Any angle between the core and axis would cause additional losses. It is equally important the cleave be as clean as possible and that there are no imperfections that would scatter light and once again, cause extra losses.

In the fabrication setup, we have access to a conventional microscope view and a white light interferometer~(WLI). The WLI in particular is essential in ensuring the quality of the cleave. As one can imagine, the quality of the fiber cleaver is of paramount importance. Good results have been achieved with both automatic and manual cleavers.

\begin{figure}[h]
    \centering
    \includegraphics[width=0.4\textwidth]{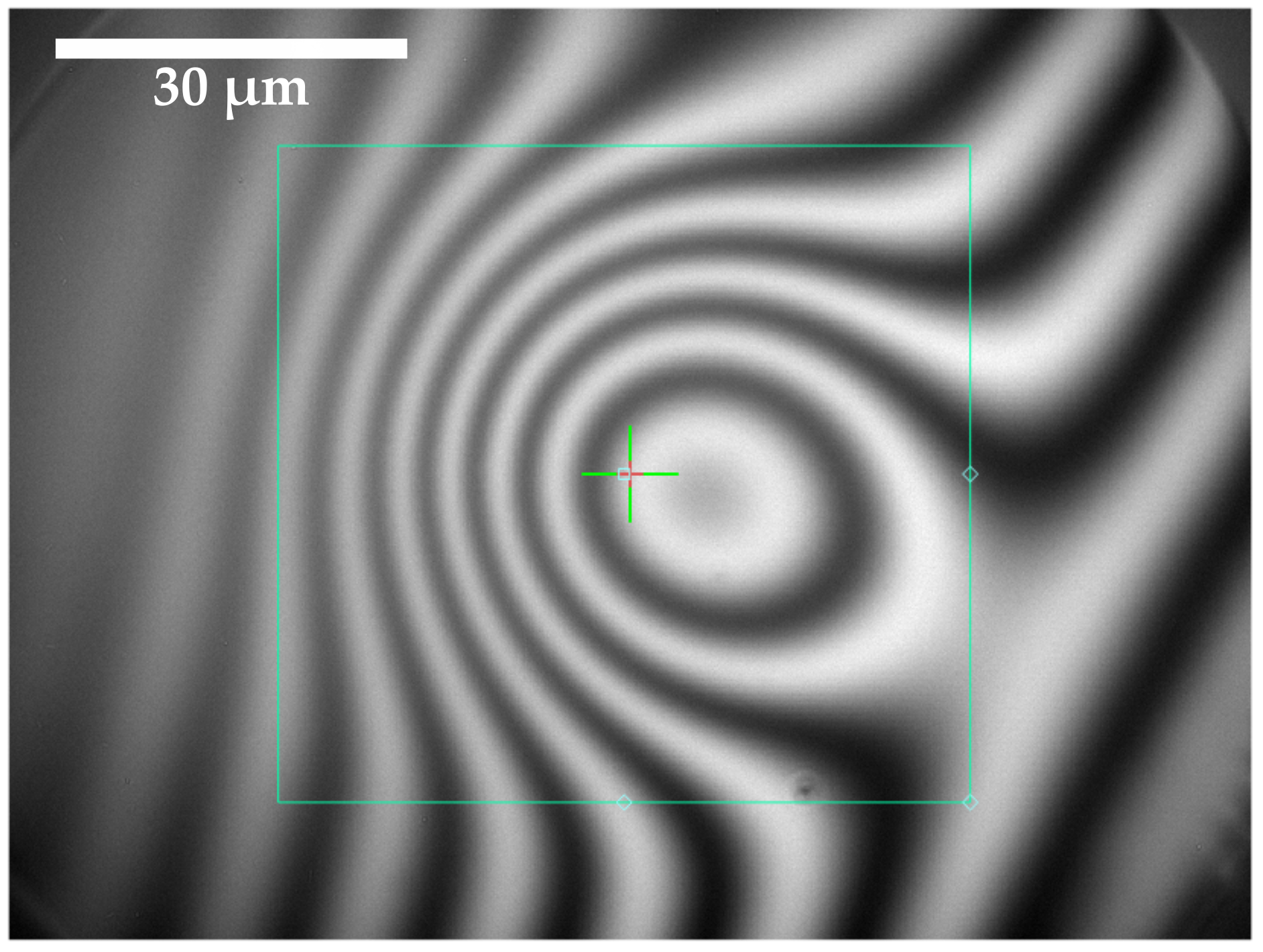}
    \caption{A white light interferometer (WLI) image of the shot fiber facet. The interference pattern indicates an angle between the imaging axis and the fiber facet and the shot indentation is identified by the rings in the center of the image. With the knowledge of the imaging wavelength, the radius of curvature of the mirror is extracted by fitting a parabolic profile to the interferometer image.}
    \label{fig:WLI}
\end{figure}

The next step
is the actual shooting. In order to build a stable cavity in practice, at least one of the mirrors needs to be curved.
Since the exoskeleton mirror is flat,
the curvature
of the fiber mirror
is created by shooting the substrate with a strong CO$_2$-laser pulse. The pulse is short enough that the fiber does not melt and form a convex shape, but rather the pulse evaporates part of the fiber in proportion to the intensity. The transverse intensity profile of the laser beam is Gaussian and the successfully fabricated fiber follows the intensity profile of the pulse. After the shot, the result is again confirmed with the WLI (Figure \ref{fig:WLI}). As the final step, the radius of curvature is fitted with a homebuilt analysis and control program. For the particular fiber used in the experiment, we found a radius of curvature of \SI{297.8}{\micro \meter}.



\subsection{Exoskeleton mirror}
\label{subsec:exo}

The exoskeleton mirror fabrication starts with the deposition of approximately \SI{200}{\nano\meter} PECVD SiO$_2$ on both sides of a \SI{775}{\micro\meter} thick <100> silicon wafer: this is done to ensure that the silicon surface remains pristine, in anticipation of the anodic bonding step to a pyrex wafer towards the end of the fabrication process. After HMDS vapour priming of the wafer, a thick ($~\SI{10}{\micro\meter}$) layer of positive photoresist (AZ4562) is spun on one of the sides of the silicon wafer and softbaked at \SI{90}{\celsius} for 5 minutes. A phononic crystal pattern is written onto the photoresist using a Heidelberg MLA 150 massless aligner and the resist is developed using AZ 726 developer. The pattern is subsequently transferred into the underlying PECVD SiO$_2$ layer using an ICP etch based on a CF$_4$ and H$_2$ gas mixture. Using water-soluble crystalbond, the patterned Si-wafer is attached to a carrier wafer and etched for approximately 1.5 hours in a deep reactive ion etching tool at \SI{0}{\celsius} using a Bosch process. This etch transfers the phononic crystal structure into the silicon substrate, with resultant vertical sidewalls. After the etch, the device wafer is removed from the carrier and rinsed thoroughly in water, to remove crystalbond residues. We then use NMP to remove the photoresist from the device wafer, followed by a 2-minute buffered hydrofluoric acid etch, stripping away the PECVD SiO$_2$ layers from both sides of the substrate. Finally, the wafer is cleaned in a hot piranha solution, alongside a thin (\SI{100}{\micro\meter}) pyrex wafer. After air drying both wafers, we use a commercial wafer bonder (SB6 from S\"{u}ss MicroTec) to anodically bond the phononic crystal structured silicon wafer and the pyrex wafer. Sputter deposition of a mirror coating on the pyrex surface concludes the fabrication of the mirrors, where the silicon phononic ``exoskeleton'' dictates the mechanical properties of the joint structure.

\section{Setup}

\subsection{Mechanical setup}

\begin{figure}[h]
    \centering
    \includegraphics[width=0.7\textwidth]{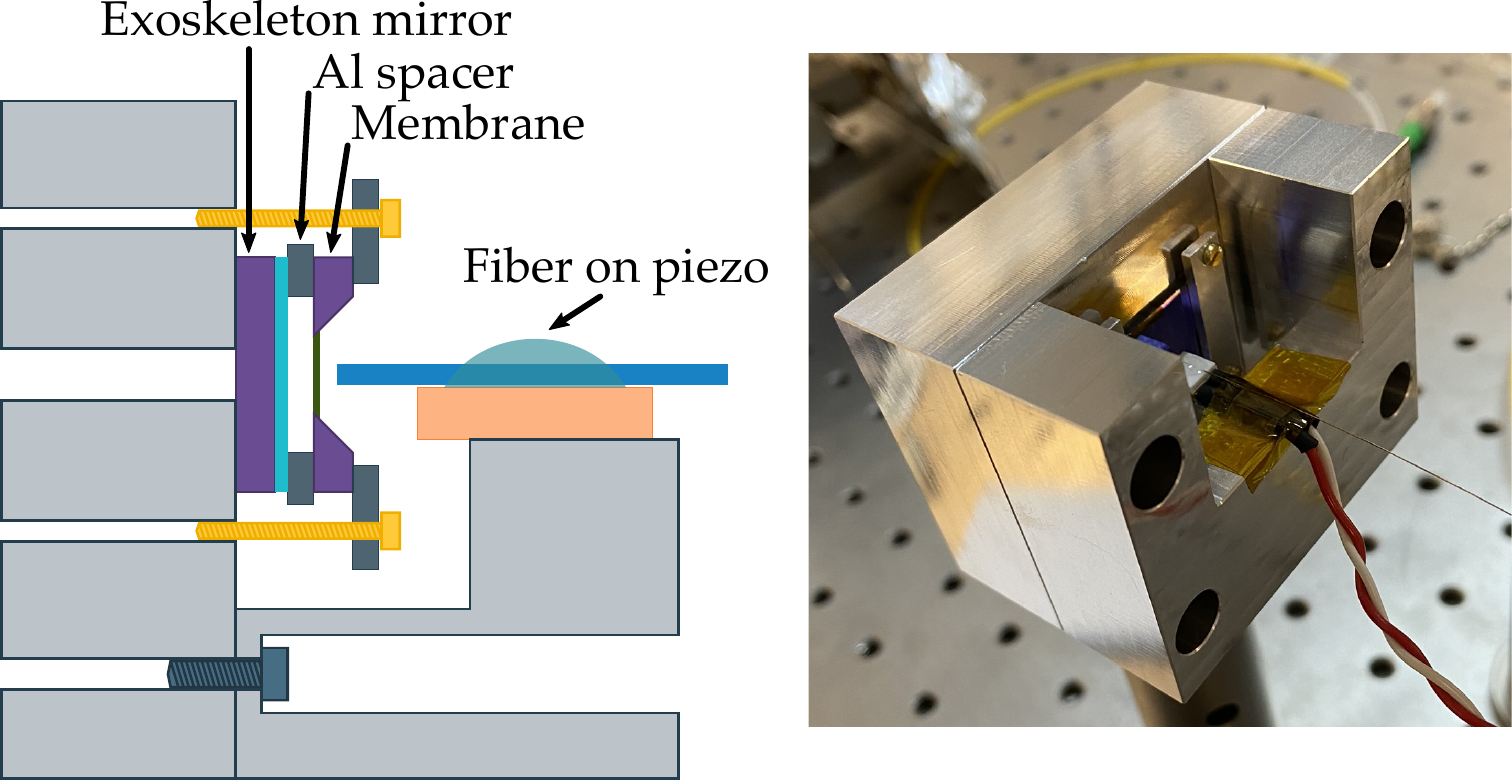}
    \caption{Schematic and photograph of the optomechanical cavity assembly. The top part of the aluminum piece holding the fiber, containing the top two screws, is omitted in the schematic for clarity.}
    \label{fig:si_cavity}
\end{figure}

The cavity design, shown in Fig.~\ref{fig:si_cavity}, was inspired by Ref.~\cite{Janitz2017}. It consists of a flat aluminum ``base'' onto which the exoskeleton mirror is clamped by a square clamp with a square hole at its center, and a second aluminum piece hosting a groove for the shear piezo onto which the fiber with the fiber mirror is glued
using
UV-curable glue. The two aluminum pieces can be screwed together via
four holes in the corners.
First, the alignment to the bare cavity is found in the following way: 1.~the exoskeleton is clamped, 2.~the two aluminum halves are screwed together and 3.~the fiber is brought in on a separate five-axis stage, gently taped onto another shear piezo on this stage (with the ``cavity'' shear piezo already fixed in the groove on the latter aluminum piece).
Once a good cavity is found with the five-axis stage, the fiber is glued onto the cavity piezo, while the stage and the other piezo are removed.
%
The alignment achieved in
this way
proved robust against taking the two aluminum halves apart and back together, and moving the exoskeleton mirror in the process. The membrane can thus be subsequently introduced atop the exoskeleton (and underneath the clamp), with strips of consumer-grade aluminum foil used as spacers.

What necessitates the use of a spacer is the fact that the silicon chip is approximately $\SI{500}{\micro\meter}$ thick, while, given the typical radii of curvature of the fiber mirrors, stable cavities can be formed with lengths up to $\sim \SI{300}{\micro\meter}$. 
The actually assembled cavities featured lengths in the range of \SIrange{50}{100}{\micro\meter}, with the membrane facing the exoskeleton mirror. 
The optical fiber is approached to the membrane through the membrane chip from the backside. This is possible since the fiber diameter is smaller than the membrane's release window in the membrane chip. We made the initial assemblies with Mylar sheets as spacers, but they proved unstable in vacuum. 
We found that strips of kitchen-grade aluminum foil met the thickness (measured to be $\SI{15.5}{\micro\meter}$ thick) and stability requirements and that cleaning them with IPA was enough to make them ultra-high vacuum (UHV) compatible.

Once clamped, the membrane pitch-yaw alignment can be fine-tuned via
four small screws on the clamp. With the fiber glued and for a given membrane pitch-yaw alignment, the cavity has no degrees of freedom as used in this work, save for the length tuning via the shear piezo.
However, the cavity design in principle allows for in situ tuning of the membrane position along the cavity axis if a ring piezo is introduced between the membrane and the clamp.

The assembly is placed inside a UHV chamber that is pumped with an ion pump to avoid vibrations.
The fiber forming the cavity is routed through a homebuilt teflon feedthrough \cite{Abraham1998} and spliced to a fiber pigtail. The chamber is evacuated to $p_\mathrm{UHV} \approx \SI{2e-8}{\milli\bar}$.

\subsection{Optical setup}

\begin{figure}[h]
    \centering
    \includegraphics[width=0.8\textwidth]{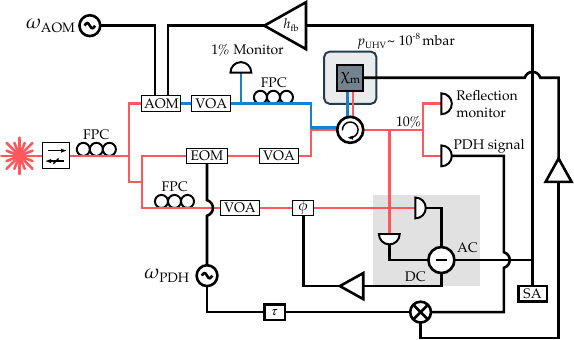}
    \caption{Schematic of the full optical setup. FPC: fiber polarization controller, AOM: acousto-optic modulator, VOA: variable optical attenuator,  PDH: Pound-Drever-Hall, SA: spectrum analyzer. The probe and local oscillator beams are shown in red and the cooling beam is shown in blue. $\tau$ indicates a phase delay used to acquire the appropriate PDH signal. $\phi$ shows a fiber stretcher for locking the homodyne setup.
    }
    \label{fig:si_setup}
\end{figure}

A schematic of the full optical setup is shown in Fig.~\ref{fig:si_setup}. We use a single laser source whose wavelength is approximately \SI{1542}{\nano\meter}. The source is protected with an external fiber-based isolator and routed trough a fiber polarization controller~(FPC).
Using a nonpolarizing beamsplitter, \SI{70}{\percent} of the light is sent to the feedback arm (i.e. the cooling beam).  
The remainder is further split into the probe beam and the local oscillator~(LO). The LO arm contains an FPC for optimizing the polarization overlap with the probe and a variable optical attenuator~(VOA) for controlling the LO power.
In the probe beam path, there is a fiber-based electro-optic modulator~(EOM) for generating the Pound-Drever-Hall~(PDH) error signal for locking the cavity on resonance and a VOA for controlling the probe strength.
%

The feedback arm is routed through a fiber-based acousto-optic modulator~(AOM) which shifts the beam frequency down by $\omega_\mathrm{AOM} = 2\pi\cdot \SI{80}{\mega\hertz}$ (and, since the probe is locked on resonance with the cavity via the PDH scheme, sets the cooling beam's detuning to $\SI{-80}{\mega\hertz}$).
The feedback control is applied via amplitude modulation around the mechanical frequency. The purpose of the frequency offset is to provide sideband cooling and stabilize the system (e.g. against thermal intermodulation noise), as well as to spectrally separate the feedback from the probe.
After the AOM, there is a VOA after which \SI{1}{\%} of the light is diverted to a power meter for monitoring the amount of power in the cooling beam. The polarization is again controlled with an FPC to minimize the polarization overlap between the probe and the cooling beam, in order to further avoid contamination of the measurement by the direct cooling beam signal. After this FPC, the cooling beam is combined with the probe (after the VOA on the probe) and the two are routed to the cavity (shown as $\chi_\mmr$) through a circulator. 

The circulator diverts the reflected light to be detected for several purposes. A total of \SI{10}{\%} of the reflected power is split into the part used to monitor the DC reflection and the part used to generate the PDH error signal. The other \SI{90}{\%} are combined with the LO in a variable fiber coupler and measured in a balanced homodyne detection scheme (indicated by the grey rectangle). 

Note that, while the cooling beam is also present in the reflection: 1. it just amounts to an immaterial offset in the detected DC power, 2. it is not detected in the PDH scheme as the signal of interest is at $\omega_\mathrm{PDH}$ and 3. its polarization is orthogonal to that of the LO and it anyway appears at $\omega_\mathrm{AOM}$ in the homodyne spectrum, as explained above. Polarization drifts are discussed in Sec.~\ref{sec:excess_noise}.
The DC part of the homodyne output is taken to the piezo of a fiber stretcher in the LO path, for controlling the relative phase between the LO and the probe (shown as $\phi$), thus locking the homodyne to the grey fringe. The AC part of the homodyne signal is taken to a spectrum analyzer~(SA) to record the mechanical spectra. At the same time, this measurement record can be filtered by a Red Pitaya, using the IQ~module of the open-source software package PyRPL~\cite{Neuhaus2017} to generate the kernel $h_\fbmr$. The filtered signal is used to modulate the amplitude of the feedback beam via the AOM, and thus provide the feedback force.

\section{Calibrations}

\subsection{Occupancy}
\label{subsec:occupancy_cal}

The data acquired on the spectrum analyzer, namely the voltage spectra $\bar S_{VV}$ given in $\si[per-mode=symbol]{\square\volt\per\hertz}$, need to be converted to apparent displacement spectra $\bar S_{yy}$, in $\si[per-mode=symbol]{\square\meter\per\hertz}$ or normalized to the zero-point fluctuations. %
Given that the experiment is conducted at room temperature, the membrane has essentially unlimited time to thermalize and one might think it can be safely taken to be at $T = \SI{300}{\kelvin}$.
There are, however, certain caveats for extracting the correct conversion factor in our experimental conditions. 
Specifically, we want the calibration to hold for a given feedback run, which typically includes a strong probe field. 
The quantum backaction from the probe effectively raises the bath temperature, which we take into account by considering the effective bath occupancy to be $\bar n_\thmr (1 + C_\qmr)$. 
Second, even a small detuning of this strong beam from cavity resonance can lead to appreciable dynamical backaction effects and, if not accounted for, to an erroneous calibration.
We thus ensure the dynamical backaction from the cooling beam already present in the system overwhelms the same effect from the probe, employing sufficient power and making use of its detuning being $\SI{80}{\mega\hertz}$ larger.

\begin{figure}[h]
    \centering
    \includegraphics[width=0.7\textwidth]
    {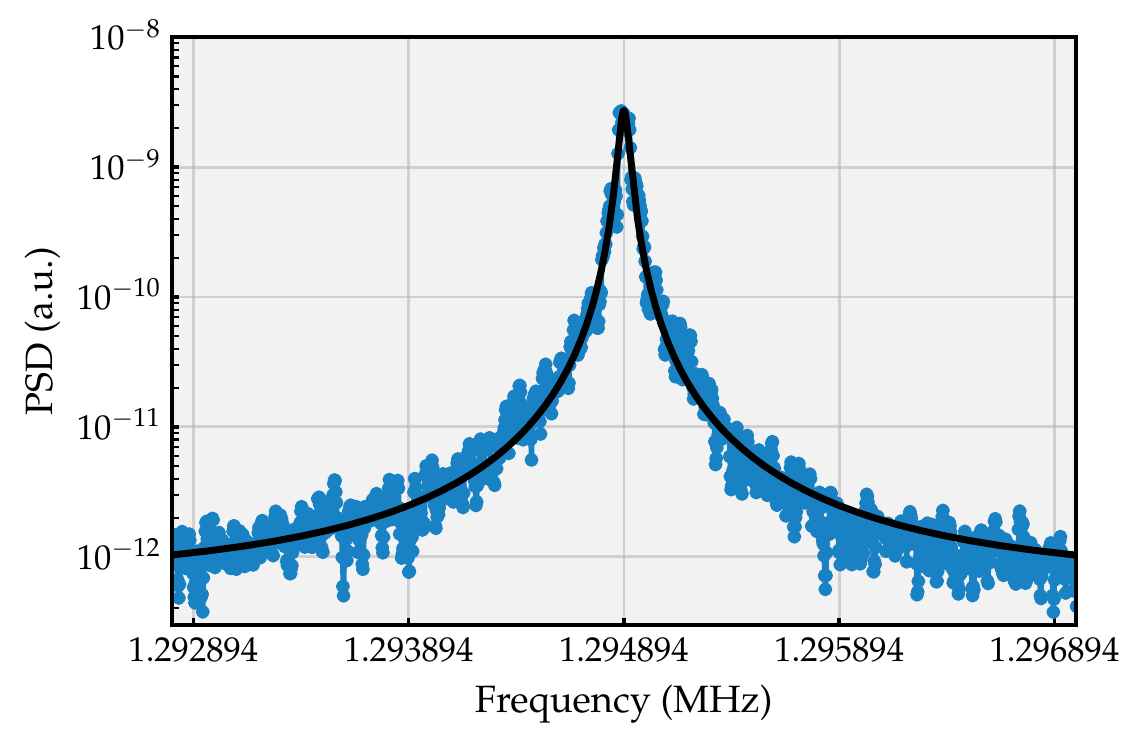}
    \caption{Mechanical anchor spectrum for calibrating the occupancy, obtained with pure sideband cooling, with the input cooling beam power set to \SI{780}{\micro\watt}. The fit to the model given by eq.~(\ref{eq:Sxx}) with $h_\fbmr = 0$, shown as a solid black line, is used to extract $\Gamma_\mmr^\tot$ (and $\Omega_\mmr^\tot$).}
    \label{fig:si_ncal}
\end{figure}

With this in mind, we apply the following procedure for the feedback run: 1. the probe power is set in advance and the PDH lock is initiated; 2a. the cooling beam is turned on and the power is set such that it dominates dynamical backaction, namely to the same value as that of the rightmost datapoint in Fig.~2; 2b. an ``anchor'' spectrum taken at this stage is used for calibration; 3. the feedback is turned on and the gain is varied. The occupation of the anchor spectrum, taken during step 2b., can be written as

\begin{equation}
   \bar n =\frac{ \bar n_\mathrm{th} \Gamma_\mmr (1+C_\mathrm{q}) + \bar n_\mathrm{m}^\mathrm{c} \Gamma_\mathrm{m}^\mathrm{c}}{\Gamma_\mathrm{m}^\mathrm{tot} } \, ,
   \label{eq:si_ncal}
\end{equation}
where
$\bar n_\mathrm{m}^\mathrm{c}$ is the minimum occupancy attainable with sideband cooling, as given in the main text.
Since the first term on the right-hand side overwhelms the quantum backaction of the cooling beam by a large margin and, at the same time, the dynamical backaction is dominated by the cooling beam, we approximate
$\bar n_\mathrm{m}^\mathrm{c} \Gamma_\mathrm{m}^\mathrm{c}/\Gamma_\mmr^\tot \approx \bar n_\mathrm{m}^\mathrm{c}$.
Note that
$\Gamma_\mmr^\tot$
is extracted from the spectrum (based on fit shown in Fig.~\ref{fig:si_ncal}), whereas all other quantities on the right-hand side can be measured independently. In particular, we need to determine the bare mechanical linewidth $\Gamma_\mmr$, which is discussed in the following subsection. 
Finding the calibration is then a simple matter of finding the constant $K$ for which
\begin{align}
  K \cdot \bar S_{VV}(\Omega)  = \bar S_{yy}(\Omega) =  \bar S_{xx}(\Omega) +   \bar S_{xx}^\impmr 
  \intertext{with}
  \int_{0}^\infty \frac{\bar S_{xx}(\Omega)}{2 x_\zpf^2} \frac{\dd\Omega}{2\pi} 
    = \bar n + \frac{1}{2} \, .
    \label{eq:si_ncal2}
\end{align}

\begin{figure}[h]
    \centering
    \includegraphics[width=0.65\textwidth]{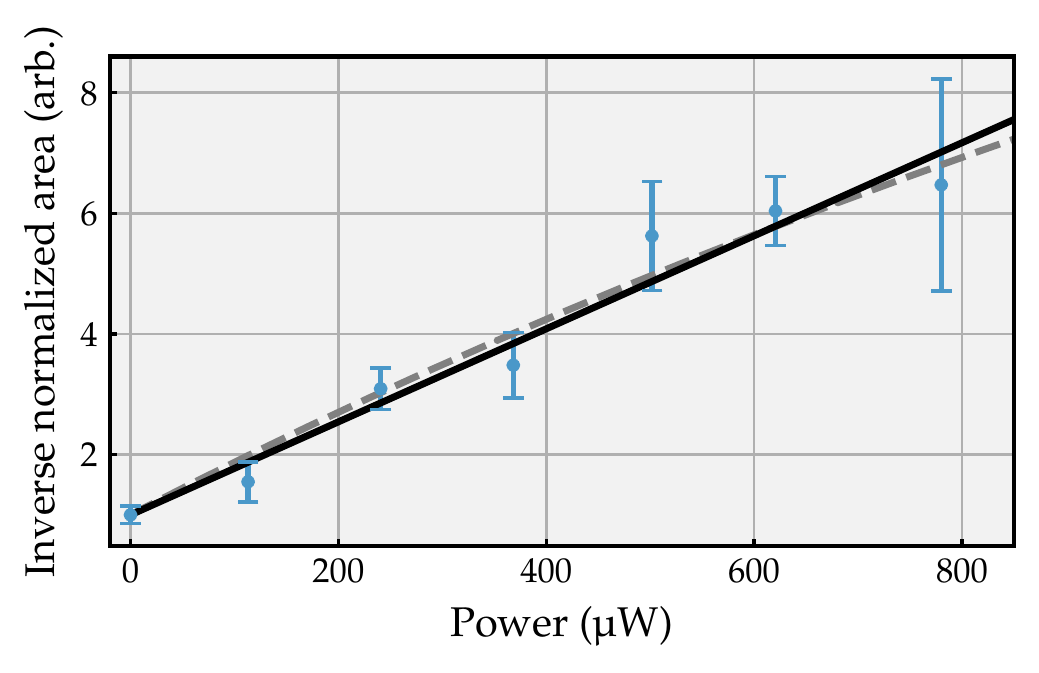}
    \caption{Inverse of the normalized mechanical area as a function of cooling beam power. The solid black line is a fit to the model of pure dynamical backaction. The model corresponding to the dashed gray line attempts to account for photothermal and intensity noise heating. The additional fitted parameter is compatible with zero classical heating.
    }
    \label{fig:si_cl_heating}
\end{figure}

It should also be stressed that classical backaction (due to photothermal and intensity noise heating) should also in principle be included in the calibration, i.e. eq.~(\ref{eq:si_ncal}). To show that this contribution is negligible, we inspect the mechanical area $A$
as a function of incident cooling beam power $P_\inmr$ (with the probe power fixed), presented in Fig.~\ref{fig:si_cl_heating}. The areas are extracted from the same dynamical backaction data for which the optical spring and optical damping rate are shown in Fig.~2 of the main text. First we make a fit to the model of pure dynamical backaction. Specifically, we take eq.~(5), neglect the quantum backaction and, since the optical damping rate induced by the cooling beam is proportional to its input power, we end up with
\begin{align}
    \frac{A}{A_0} = \frac{\Gamma_0}{a_\mathrm{dba} P_\inmr + \Gamma_0} \nonumber
\end{align}
in this model.
Here, $A_0$ and $\Gamma_0$ are the area and linewidth extracted for the first datapoint (from a fit to $|\chi^\tot_\mmr|^2$), and they are fixed, wheres $a_\mathrm{dba}$ is a free parameter. We choose to fit the inverse of this normalized area, such that we have
\begin{align}
    \left(\frac{A}{A_0}\right)^{-1} = \frac{a_\mathrm{dba} P_\inmr}{\Gamma_0} + 1
\end{align}
for the solid black line. Instead, the dashed gray line includes classical heating (by either photothermal or intensity noise heating, including TIN-related intensity noise) as 
\begin{align}
    \left(\frac{A}{A_0}\right)^{-1} = \frac{a_\mathrm{dba} P_\inmr + \Gamma_0}{(1 + a_\mathrm{eh} P_\inmr) \, \Gamma_0} \, , \nonumber 
\end{align}
with $a_\mathrm{dba}$ and $a_\mathrm{eh}$ as fitted parameters. A fit to this generalized expression yields
$a_\mathrm{eh} = \SI[separate-uncertainty = true]{2(3)e-8}{\per\micro\watt}$,
showing that the data is compatible with zero classical heating. Note that the incident cooling beam power for the feedback experiment was $P_\inmr = \SI{780}{\micro\watt}$.
While we do not have an equivalent measurement for the probe beam, we emphasize that the same probe power used in the feedback run (corresponding to $C_\qmr = 0.1$) is present also in this measurement of dynamical backaction. This suggests that both photothermal and intensity noise heating are negligible in the current experiment. For completeness, we note that the input probe power in question is $P_\mathrm{probe} \approx \SI{520}{\micro\watt}$. Accounting for the different cavity filtering due to the two beams’ different detunings, the corresponding intracavity probe power is still smaller than for the highest cooling beam powers shown in the figure above. Specifically, this probe power is equivalent to \SI{640}{\micro\watt} in the graph.

\subsection{Bare mechanical damping rate}

For soft-clamped membrane resonators, the bare mechanical damping rate is only a few millihertz and below the bandwidth of most spectrum analyzers. Thus, the damping rate cannot be accurately estimated based on the spectral response alone. Instead, the membrane
is placed in a purpose-built interferometer setup. The membrane motion is excited with a piezo and once the excitation is turned off, the amplitude of the mechanical mode decays exponentially at a rate $\Gamma_\mmr$. This exponential decay is then fitted to the measured time trace of the mechanical amplitude.

\begin{figure}[h]
    \centering
    \includegraphics[width=0.7\textwidth]{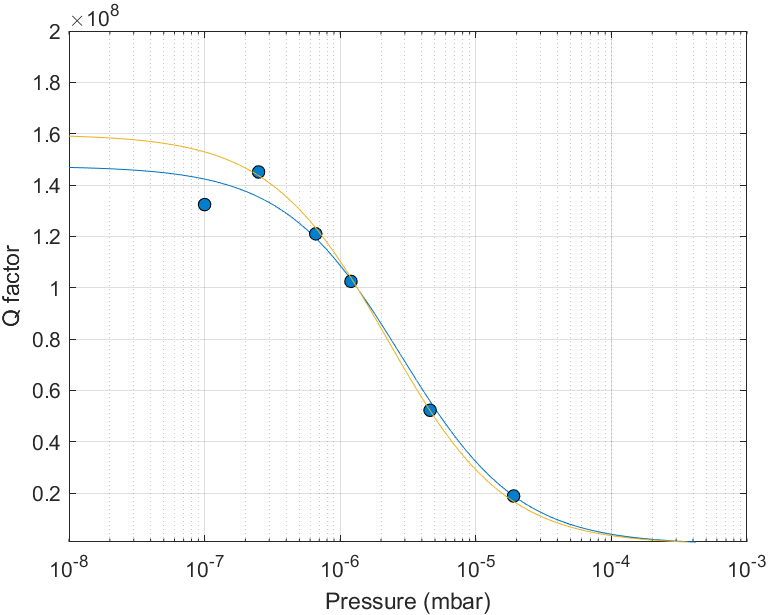}
    \caption{$Q$-factor of a \SI{17}{\nano\meter}-thick Dandelion as a function of pressure. The yellow and blue lines are fits to the data  excluding and including the point at lowest pressure, respectively. This point was acquired in a separate measurement, after baking the chamber with the membrane inside, in an attempt to reduce the pressure further. It appears this has lowered the Q-factor of the mechanical mode. We make use of the yellow fit to correct for the pressure difference between the interferometer and main experiment's vacuum conditions.}
    \label{fig:si_qvp}
\end{figure}

However, the interferometer vacuum chamber only reaches a pressure of approximately $p_\mathrm{intf} \approx \SI{2e-7}{\milli\bar}$, an order magnitude higher than the UHV chamber used in the feedback experiment. Gas damping is still significant at this pressure and consequently the quality factor of the mechanical resonator device should increase in the UHV chamber compared to the value found in the interferometer measurement. The increase can be estimated by measuring the decay rate at different pressure points and fitting the total quality factor
\begin{align}
Q^{-1}(p) = Q_0^{-1} + a_\mathrm{Q} \cdot Q_D^{-1}(p) \, , \nonumber
\end{align}
where $Q_0$ is the mechanical quality factor without effects of gas damping (i.e. at zero pressure) and
\begin{align}
  Q_D(p) = \frac{\rho \, h \, \Omega_\mmr}{4} \sqrt{\frac{\pi}{2}} \sqrt{\frac{R T}{M}} \frac{1}{p} \, , \nonumber
\end{align}
with $R$ the molar gas constant, $\rho = \SI{3170}{\kilo\gram\per\meter\cubed}$ the mass density of silicon nitride, $M = \SI{28.97}{\gram\per\mol}$ the molar mass of air and $h$ the membrane thickness. The fit for a Dandelion membrane is shown in Fig.~\ref{fig:si_qvp}, with the two fitting parameters being $Q_0$ and
an experimental correction factor
$a_Q$. While the design is not the exact one used in this work, the thickness is essentially the same. We therefore assume the gas damping characteristics to be similar and estimate that, for the membrane used in the experiment, the reduced gas damping increases the bare mechanical quality factor by approximately \SI{10}{\percent} compared to the value found in the interferometer characterization.

On the other hand, after finishing the feedback measurements with the optomechanical cavity, we took the cavity apart and measured the membrane $Q$-factor in the interferometer chamber once again. At nominally the same pressure as
in the prior interferometer measurement,
we have seen a reduction of about \SI{20}{\percent}. The bare mechanical linewidth $\Gamma_\mmr$, used in the calibration according to eq.~(\ref{eq:si_ncal}), is derived starting from this reduced value by taking into account also the reduced gas damping.

\section{Fiber cavity reflection model}
\label{sec:fibcav_refl}
The reflection of an optical cavity comprising a fiber mirror as the input coupler has a particular lineshape, which arises as a consequence of the spatial mode filtering of the reflected light done by the fiber. We describe this lineshape, shown in Fig.~1d) of the main text, using a model derived in~\cite{Gallego2016}. For convenience, we provide here a shorter version of said derivation, using essentially the same notation.

Since spatial mode filtering is of the essence, it is instructive to write all the fields $E_i = \mathcal E_i \ket{\psi_i}$ as a product of $\mathcal E_i$ containing the complex amplitude and time dependence, and $\ket{\psi_i}$ containing the spatial mode components. For the more general case of a fiber cavity, the relevant modes are the forward- and backward-propagating modes in the fiber, $\ket{\psi_\fmr^\pm}$, the forward- and backward-propagating cavity modes, $\ket{\psi_\cav^\pm}$, and the spatial mode of the field promptly reflected at the input mirror, $\ket{\psi_\rmr}$. Their respective overlap integrals are as follows:
\begin{align}
    &\braket{\psi_\cav^+|\psi_\fmr^+} \equiv \alpha \, , \nonumber \\
    &\braket{\psi_\cav^-|\psi_\rmr} = \braket{\psi_\cav^+|R^\dagger R|\psi_\fmr^+} = \alpha \, , \nonumber \\
    &\braket{\psi_\cav^-|\psi_\fmr^-} = (\braket{\psi_\cav^+|\psi_\fmr^+})^* = \alpha^* \, , \nonumber \\
    &\braket{\psi_\fmr^-|\psi_\rmr} \equiv \beta \, . \nonumber
\end{align}
Here we have used the fact that reflection is a unitary transformation ($R^\dagger R = \mathbb{1}$) and that overlap amplitudes are conjugated when the directions of propagation of both modes are changed~\cite{Hood2001}.

The input field can now be written as\footnote{To be consistent with Sec.~\ref{sec:tin}, we should write it as $E_{\inmr,1}$, but we have decided to drop the subscript here for brevity.} $E_\inmr = \mathcal E_\inmr \ket{\psi_\fmr^+}$, and the promptly reflected field as $E_\rmr = r_1 \mathcal E_\inmr \ket{\psi_\rmr} \approx \mathcal{E}_\inmr \ket{\psi_\rmr}$. $r_i$ is the amplitude reflection coefficient of the $i$-th mirror. The cavity leakage field can further be shown to be
\begin{align}
E_\mathrm{leak}(\Delta) &= -\mathcal E_\inmr \zeta(\Delta) \braket{\psi_\cav^+|\psi_\fmr^+} \ket{\psi_\cav^-} \, , \nonumber
\end{align}
with
\begin{align}
\zeta(\Delta) = \frac{t_1^2 r_2 e^{i \phi_\cav}}{1 - r_1 r_2 e^{i \phi_\cav}} \, . \nonumber
\end{align}
Here $t_i$ is the amplitude transmission coefficient for the $i$-th mirror and $\phi_\cav = 2 \pi \Delta/\omega_\mathrm{FSR}$ is the cavity round-trip phase, with $\omega_\mathrm{FSR}$ the cavity free spectral range in angular frequency units.

The reflected field can be decomposed into a contribution which is mode-matched to the cavity, and that which is not. The former therefore interferes with the leakage field. For a free-space cavity, doing this decomposition makes it straightforward to find the total reflected power (normalized to the input) as $P_\outmr(\Delta)/P_\inmr = \varepsilon \left| 1 - \zeta(\Delta) \right|^2 + (1 - \varepsilon)$, with $\varepsilon = |\alpha|^2$ the mode-matching efficiency to the cavity mode. A Taylor expansion of $\zeta(\Delta)$ around resonance then leads to the familiar Lorentzian dip, the depth of which is proportional to, and ultimately limited by, $\varepsilon$.

For a fiber cavity, the total reflected field is again the sum of the promptly reflected one and the leakage from the cavity into the reflection port, but the light needs to couple back to the fiber mode in order to reach the detector. The guided field is found by projecting the reflected field onto the fiber mode,
\begin{align}
    E_\outmr(\Delta) = ( \bra{\psi_\fmr^-} E_\rmr + \bra{\psi_\fmr^-} E_\mathrm{leak}(\Delta) ) \ket{\psi_\fmr^-} = \mathcal E_\inmr ( \braket{\psi_\fmr^-|\psi_\rmr} - \zeta(\Delta) \braket{\psi_\cav^+|\psi_\fmr^+} \braket{\psi_\fmr^-|\psi_\cav^-} ) \ket{\psi_\fmr^-} \, . \nonumber
\end{align}
It follows that the detected power is
\begin{align}
    P_\outmr = E_\outmr^\dagger E_\outmr = P_\inmr \left| \beta - \alpha^2 \zeta(\Delta) \right|^2 = \frac{P_\inmr}{\varepsilon^2} \left| \beta (\alpha^2)^* - \varepsilon^2 \zeta(\Delta) \right|^2 \, . \nonumber
\end{align}
Again doing a Taylor expansion around resonance yields
\begin{align}
    \frac{P_\outmr(\upsilon)}{P_\inmr} = \eta_\rmr - \eta_\mathcal{L} \left( \frac{1}{1 + \upsilon^2} - \mathcal{A}\frac{\upsilon}{1 + \upsilon^2} \right) \, ,
\label{eq:si_refdip}
\end{align}
with $\upsilon = 2\Delta/\kappa$. This is an asymmetric dip consisting of a Lorentzian of amplitude
\begin{align}
    \eta_\mathcal{L} = \frac{4 T_1}{L_\tot} \left( \RE[\beta (\alpha^2)^*] - \varepsilon^2 \frac{T_1}{L_\tot} \right) \nonumber
\end{align}
and its corresponding dispersive lineshape with relative amplitude
\begin{align}
    \mathcal A = \frac{\IM[\beta (\alpha^2)^*]}{\RE[\beta (\alpha^2)^*] - \varepsilon^2 \frac{T_1}{L_\tot}} \, , \nonumber
\end{align}
with $L_\tot = T_1 + T_2 + l_1 + l_2$, $T_i = |t_i|^2$ and $l_i$ the power transmissivities and other (absorption and scattering) power losses of the $i$-th mirror respectively, and $\RE[\dots]$ and $\IM[\dots]$ denoting the real and imaginary parts. The off-resonant power loss to the fiber cladding modes, resulting in a reduced off-resonance level $\eta_\rmr = |\beta|^2$, can be shown to be a consequence of the fiber mirror decentration alone.

We can see that in the fiber cavity case the dip depth is no longer limited by $\varepsilon$. Furthermore, to extract $\varepsilon$ from the reflection, it is not enough to know the cavity
overcoupling
$\eta_\mathrm{c}$. Rather, one requires the knowledge of the overlap integrals, which in turn depend on the cavity mirror parameters (radii of curvature, fiber mirror decentration(s) and fiber mode field radius) and cavity alignment parameters (cavity length and the pitch and yaw alignment angles)~\cite{Gallego2016}. We do not possess the knowledge of all the parameters, in particular the alignment angles, to sufficient precision. Instead, we estimate the mode-matching efficiency $\varepsilon$ from the cavity transmission, where there is no spatial filtering by the fiber mode. To extract the linewidth $\kappa$, the reflection dip shown in Fig.~1d) of the main text is fitted to eq.~(\ref{eq:si_refdip}) with $\kappa$, $\eta_\mathcal{L}$ and $\mathcal{A}$ as free parameters. The off-resonance level $\eta_\rmr$ is fixed to a value measured independently, with a retro-reflector (cf. Sec.~\ref{sec:meas_loss}).

\section{Cavity frequency noise}
For the measurement of the cavity frequency noise, shown in Fig.~3c) of the main text, there was no membrane inside the cavity. The calibration of the noise is done in a way inspired by Gorodetsky et al.~\cite{Gorodetsky2010}, wherein an external phase modulation of known amplitude $\varphi_0$ (and at frequency $\Omega_\modmr$) is used to calibrate the spectrum. Namely, the spectrum can then be calibrated via
\begin{align}
\int \bar S_{VV}(\Omega_\modmr) \dd \Omega = C \cdot \int \bar S_{\varphi\varphi}(\Omega_\modmr) \dd \Omega = C \cdot \varphi_0^2 \, ,
\end{align}
where we used the fact that the phase spectrum is $\bar S_{\varphi\varphi}(\Omega_\modmr) = \delta(\Omega_\modmr - \Omega) \varphi_0^2$.
However, the crux of the Gorodetsky method is that the signal and the local oscillator for the phase-sensitive detection are both subjected to this external modulation. This ensures that any signal at $\Omega_\modmr$ when the cavity is locked arises from phase-modulation done by the cavity itself, i.e. that the external tone undergoes exactly the same transduction as the mechanical sidebands generated inside the cavity. Because of large residual amplitude modulation from the phase modulator, we found it challenging to operate in this regime. Instead, we chose to modulate just the signal arm with this additional, calibration tone at $\Omega_\modmr/2\pi = \SI{2.2}{\mega\hertz}$ (not shown in Fig.~3c) of the main text). The modulation depth and frequency are chosen such that the tone is clearly visible in the spectrum, while not perturbing the spectrum around the mechanical frequency $\Omega_\mmr/2\pi = \SI{1.3}{\mega\hertz}$, i.e. in the wider window where the exoskeleton bandgap is, in principle, expected.

Setting up the measurement as we did means that we can still calibrate the spectrum as explained above, but we need additional knowledge of our cavity parameters compared to the actual Gorodetsky method, since the external tone is subjected to a transduction different from that of the mechanical modulation. It can be shown that the ratio of the locked and unlocked spectral density at $\Omega_\modmr$ is
\begin{align}
    \frac{S_\mathrm{lock}(\Omega_\modmr)}{S_\mathrm{unlock}(\Omega_\modmr)} = (1 - 2\Lambda)^2 \, ,
\label{eq:si_sp_ratio}
\end{align}
with $\Lambda$ a function of the cavity
overcoupling, the mode-matching efficeincy to the cavity mode and fiber recoupling efficiency. On the other hand, for a resonant readout and in the limit $\kappa \gg \Omega$, some mechanical phase modulation $\varphi_\mmr$ is transduced as
\begin{align}
    \bar S_{\varphi\varphi}(\Omega) = \frac{64 \Lambda^2 \Omega^2}{\kappa^2}\left( \frac{\varphi_\mmr}{2} \right)^2 \, . \nonumber
\end{align}
From the ratio in eq.~(\ref{eq:si_sp_ratio}), it is possible to extract two values for $\Lambda$. The cavity phase noise for either $\Lambda$ is then straightforwardly converted to frequency noise, or displacement noise, via the frequency pull parameter $\dd\omega_\mathrm{c}/\dd x$. The cavity noise shown in the main text is calibrated with the less favourable value of $\Lambda$, i.e. the one which leads to a slightly larger residual occupation. 

\section{Imprecision thermal intermodulation noise}
\label{sec:tin}
Neglecting vacuum fluctuations, the equation of motion for the intracavity field amplitude $a$ reads
\begin{equation}
   \dot a(t) = \left( i\Delta (t) - \frac{\kappa}{2}\right)a(t) + \sqrt{\kappa_1} s_{\inmr,1} \, , \nonumber
\end{equation}
where the subscript $1$ indicates the port to which the laser is coupled, such that $s_{\inmr,1}$ is the coherent drive amplitude and $\kappa_1 = \eta \kappa$ is the cavity decay rate associated with this port. The detuning $\Delta(t)$ is modulated by the cavity frequency noise. In the fast cavity limit, where $1/\kappa$ is much smaller than the timescale on which $\Delta$ changes,
it follows that
\begin{equation}
a(t) = 2\sqrt{\frac{\eta}{\kappa}}\mcl(\upsilon(t)) s_{\inmr,1} \, ,
\end{equation}
with $\upsilon = 2\Delta/\kappa$ (as defined in Sec.~\ref{sec:fibcav_refl}) and 
\begin{equation}
\mcl(\upsilon) = \frac{1}{1-i\upsilon} \, .
\label{eq:SI_L}
\end{equation}
Expanding eq.~(\ref{eq:SI_L}) in small fluctuations $\delta\upsilon$ around the mean value $\upsilon_0$, we have
\begin{equation}
a = 2 \sqrt{\frac{\eta}{\kappa}} \mcl(\upsilon_0)(1 + i \mcl(\upsilon_0)\delta\upsilon - \mcl(\upsilon_0)^2\delta\upsilon^2 - i \mcl(\upsilon_0)^3\delta\upsilon^3 +\mathcal{O}(\delta\upsilon^4))s_{\inmr,1} \, .
\label{eq:SI_exp}
\end{equation}

According to the input-output relations, the field detected in the reflection port is
\begin{equation}
s_{\outmr,1}(t) = s_{\inmr,1} - \sqrt{\kappa_1} a(t) \, , \nonumber
\end{equation}
and for homodyne detection locked to the gray-fringe, the signal is proportional to the phase of the field
\begin{equation}
Y_{\outmr,1}(t) \propto \frac{1}{i}(s_{\outmr,1}^* - s_{\outmr,1}) \, . \nonumber
\end{equation}
In particular, the contribution from the cavity field is
\begin{equation}
Y(t) \propto \frac{1}{i}(a^* - a) \, , \nonumber
\end{equation}
for which, using eq.~(\ref{eq:SI_exp}), we have explicitly
\begin{equation}
    \frac{1}{i}(a^* - a) \propto -\frac{2}{1+\upsilon_0^2} \left( \upsilon_0 + \frac{1 - \upsilon_0^2}{1 + \upsilon_0^2} \delta\upsilon + \frac{\upsilon_0 (\upsilon_0^2 - 3)}{(1 + \upsilon_0^2)^2} \delta\upsilon^2 + \frac{\upsilon_0^4 - 6\upsilon_0^2 + 1}{(1 + \upsilon_0^2)^3} \delta\upsilon^3 + \mathcal{O}(\delta\upsilon^4)\right) \, . \nonumber
\end{equation}
It is now apparent that, contrary to the resonant measurement of the intensity ($\propto |s_{\outmr,1}|^2$, considered in~
\cite{Fedorov2020}),
it is the even-power terms that vanish for a resonant readout ($\upsilon_0 = 0$), while the odd-power terms do not (and they can in fact easily be shown to be maximal). Particularly, the linearly transduced fluctuations are detected with maximum sensitivity. While the TIN can in principle still become visible, it is further suppressed compared to the case of intensity detection due to the fact the leading nonlinear term is $\delta\upsilon^3$.

More specifically, the leading nonlinear term in the spectra is the one containing the correlation function $\langle \delta \upsilon(t) \delta \upsilon(t+\tau)^3\rangle$. For a Gaussian noise, this can be expressed using Wick's theorem~\cite[Sec.~2.8.1]{Gardiner1985} as
\begin{equation}
    \langle \delta \upsilon(t) \delta \upsilon(t+\tau)^3\rangle = 3 \langle \delta \upsilon(t)^2 \rangle  \langle \delta \upsilon(t) \delta \upsilon(t + \tau)\rangle \, ,
    \label{eq:SI_corr1}
\end{equation}
where we have also assumed that the noise is stationary, both of which are valid assumptions for typical thermal noises. In the context of mechanical TIN, it is worth noting that this term does not contain peaks at the sums and differences of mechanical resonance frequencies. The lowest-order correlation function that does is $\langle \delta \upsilon(t)^3 \delta \upsilon(t+\tau)^3\rangle$, which
contains the
Wick contraction
$\langle \delta \upsilon(t) \delta \upsilon(t+\tau)\rangle^3$. The spectral contribution of this term is
\begin{equation}
\int_{-\infty}^\infty \langle \delta \upsilon(t) \delta \upsilon(t+\tau)\rangle^3 e^{i\omega \tau} \dd \tau \propto \int_{-\infty}^\infty \int_{-\infty}^\infty \dd\omega' \dd\omega'' S_{\upsilon\upsilon}(\omega - \omega') S_{\upsilon\upsilon}(\omega' - \omega'') S_{\upsilon\upsilon}(\omega'')\, ,
\label{eq:SI_corr2}
\end{equation}
where $S_{\upsilon\upsilon}$ is the spectrum of linearly transduced fluctuations, as routinely employed throughout this document and the main text.

It is finally instructive to look at the scaling of these terms. The frequency fluctuations of a MIM cavity are dominated by the membrane displacement $\delta x$ and can therefore be expressed as $\delta \upsilon(t) = 2(g_0/\kappa) \, \delta x(t)/x_\zpf$. They are largest at RT, i.e. without any cooling of the motion, in which case $\delta \upsilon_\thmr^\mathrm{RMS} (t) = 2(g_0/\kappa) \, \sqrt{\bar n_\thmr}$. From this it follows that the spectral contribution corresponding to the correlation function~(\ref{eq:SI_corr1})
is suppressed as $\simeq (g_0/\kappa)^2 \bar n_\thmr S_{\upsilon\upsilon}$, with $(g_0/\kappa)^2 \bar n_\thmr \simeq \num{e-4}$ in our system.
This scaling is
the same as the leading nonlinear term in the case of an intensity measurement for $\upsilon_0 = 0$, but without mixing the mechanical resonances. On the other hand, the higher-order contribution~(\ref{eq:SI_corr2}), while featuring mode-mixing, scales more favorably, as
$\simeq (g_0/\kappa)^4 \bar n_\thmr^2 S_{\upsilon\upsilon}$.
Note also that we employ substantial sideband cooling of all membrane modes in a large frequency range around the mode of interest, so that the relevant cavity frequency fluctuations stemming from the membrane motion are smaller than $\delta \upsilon_\thmr^\mathrm{RMS}$. Therefore, we expect to not be limited by TIN in detection. Moreover, sideband cooling also addresses TIN-related intensity noise which, while not precluded by the above considerations, neither appears in the cavity scans (Fig.~1d) of the main text), in contrast to Fig.~2a in~\cite{Fedorov2020}, nor does it affect the bath temperature (see Subsec.~\ref{subsec:occupancy_cal}).

\section{Excess imprecision noise}
\label{sec:excess_noise}

\begin{figure}[ht]
    \centering
    \includegraphics[width=0.7\textwidth]{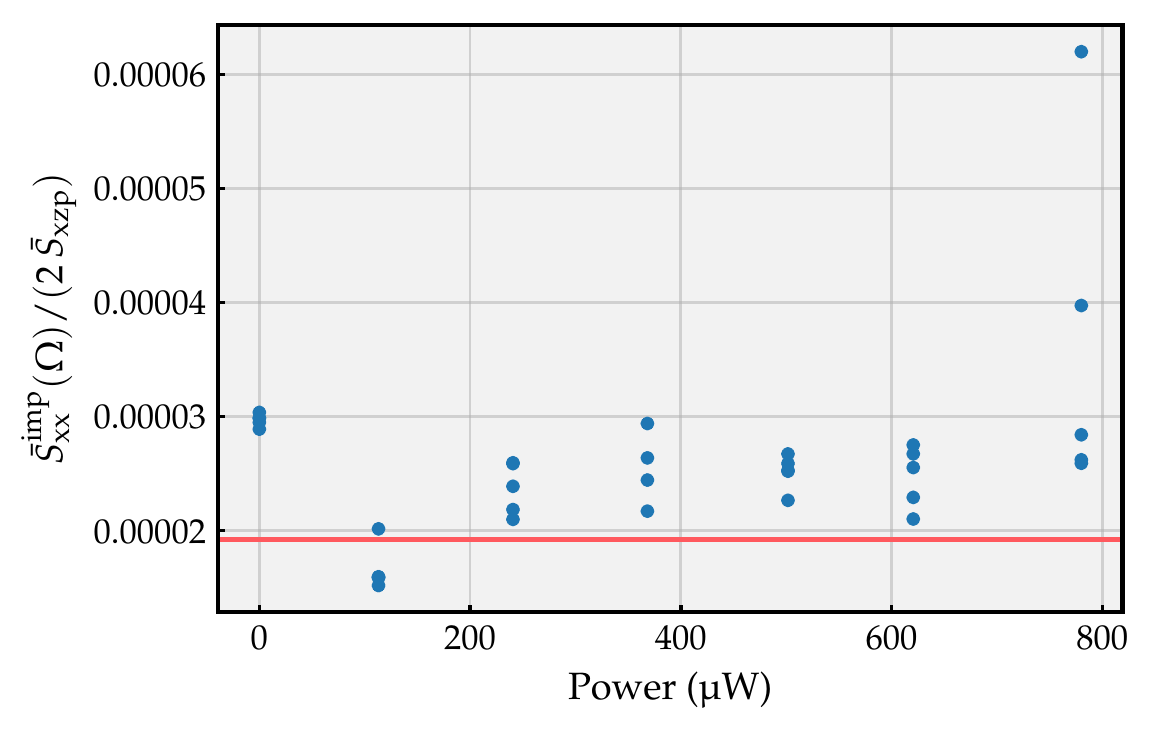}
    \caption{Imprecision level extracted from the fits to the dynamical backaction spectra used in Fig.~2 and Fig.~\ref{fig:si_cl_heating}, as a function of cooling beam power. The red line corresponds to the average of the shot-noise trace in Fig.~4b).}
    \label{fig:si_impvpower}
\end{figure}

The feedback-cooled spectra show excess classical imprecision noise, which is comparable to the shot noise, and becomes discernible in a detailed analysis (cf. Fig.~\ref{fig:si_fitted_params}a) and Fig.~4b)). In an effort to identify the source of this noise, Fig.~\ref{fig:si_impvpower} presents the fitted imprecision (with zero feedback gain) as a function of cooling beam power, as extracted from the dynamical backaction spectra.  

Note that this measurement is not equivalent to sweeping the probe power, because the cooling beam polarization is engineered to be orthogonal to that of the probe and homodyne LO. Furthermore, because the dynamical backaction spectra were intended to yield only the optical damping and spring, it is possible that these fits overestimate the imprecision, as the acquisition span may not have been sufficiently large. That said, we point out that the extracted level is above the mean shot noise level in Fig.~4a) (here represented by the red line), and that there is a large spread in the data points for the largest cooling beam power (which is the one used during the feedback run).
There is no clear dependence of the extracted noise level on the optical power (at least if the points for the largest power are excluded). Because of this, the result is inconclusive with respect to the source of the excess noise. The cavity and laser frequency noise are respectively one and two orders of magnitude smaller than the shot noise. If TIN were the cause, we would expect the noise in Fig.~\ref{fig:si_impvpower} to decrease with cooling power. This is related to the effect that cooling mechanical modes to smaller amplitudes strongly reduces their nonlinear transduction, as observed in our system before, and reported e.g. in~\cite[Sec.~5.1]{RossiThesis2020}. We would also expect the imprecision for the feedback-cooled spectra (Fig.~\ref{fig:si_fitted_params}a)) to increase with gain, on account of the feedback exciting modes with frequencies where the feedback phase is wrong. Finally, the cooling beam can appear in the spectrum due to imperfect polarization control, but Fig.~\ref{fig:si_impvpower} does not show the expected linear increase of imprecision with cooling beam power.
Therefore, yet another unknown mechanism for the excess imprecision cannot be ruled out, and should be investigated further.

The largest contribution to the noise in detection is still the quantum imprecision, which is largely due to poor detection efficiency and, in turn, cavity mode-matching. Improving the cavity mode-matching would mitigate the two outlined potential causes of classical imprecision. It would allow for a larger ratio of the LO to cooling beam power, which would reduce the effect of imperfect polarization control. On the other hand, it would also enable stronger sideband cooling, thus reducing the potential residual effect of TIN (while maintaining enough power in the LO to not be compromised by the cooling beam). 

\section{Measurements of losses}
\label{sec:meas_loss}
The cavity
overcoupling
$\eta_\mathrm{c}$ is inferred from the measurement of the power transmissivities of the fiber mirror and the exoskeleton mirror (neglecting absorption and scattering losses). The measurement of the power transmitted by the cavity $P_\mathrm{t}$ then reveals the mode-matching efficiency to the cavity mode (with the membrane inside the cavity), $\varepsilon$, since
\begin{align}
    \frac{P_\mathrm{t}(\Delta = 0)}{P_\inmr} = 4 \, \varepsilon \, \eta_\mathrm{c}(1 - \eta_\mathrm{c}) \, . \nonumber
\end{align}

We also measure the losses associated to the decentration of the fiber mirror, mostly for the purpose of estimating the total detection efficiency, but also to constrain the fit to the cavity reflection dip (cf. Sec.~\ref{sec:fibcav_refl}). Specifically, we measure the power reflected from the fiber mirror (i.e. from the cavity when the beam is far from resonance), $P_\rmr(\Delta \gg \kappa)$, and the power reflected from a retro-reflector, $P_\mathrm{rr}$, using the same input power. Then we infer 
\begin{align}
    \eta_\rmr = |\beta|^2 = \frac{P_\rmr(\Delta\gg \kappa)}{P_\mathrm{rr}} \, . \nonumber
\end{align}

To measure and optimize the homodyne interferometer visibility $\mathcal{V}$, we set the powers of the signal and LO arms to the same value and block one port of the balanced detector. The maximum and minimum measured voltages then relate to the visibility as
\begin{align}
    \mathcal{V} = \frac{V_\maxmr - V_\minmr}{V_\maxmr + V_\minmr} \, , \nonumber
\end{align}
where $V_\maxmr$ and $V_\minmr$ are the maximum and minimum voltages, respectively.

The last thing that figures into our overall detection efficiency is the quantum efficiency of the detector, $P_\mathrm{QE}$. We use the manufacturer number, $P_\mathrm{QE} = 0.8$. That, and the measured values for the other losses are listed in Table~\ref{tab:losses}.

We estimate the total detection efficiency as 
$\eta_\mathrm{det} = \varepsilon \cdot \eta_\mathrm{c} \cdot |\beta| \cdot \mathcal{V} \cdot P_\mathrm{QE} \approx \SI{1.2}{\percent}$.

\begin{table}[h!]
    \centering
    \begin{tabular}{|c|c|c|}
    \hline
    quantity & symbol & value \\
    \hline
    \hline
    cavity mode-matching efficiency & $\varepsilon$ & $\approx 0.04$ \\
    cavity overcoupling & $\eta_\mathrm{c}$ & 0.9 \\
    fiber losses & $\eta_\rmr$ & 0.42 \\
    homodyne visibility & $\mathcal{V}$ & 0.9 \\
    quantum efficiency & $P_\mathrm{QE}$ & 0.8 \\
    \hline
    \end{tabular}
    \caption{Individual contributions to the total detection efficiency.}
    \label{tab:losses}
\end{table}

\bibliography{si_refs}